\documentclass[12pt]{article}
\usepackage{epsf} 
\usepackage{float} 
\begin{document}
\baselineskip 14pt plus 2pt minus 2pt
\newcommand{\beq}{\begin{equation}}
\newcommand{\eeq}{\end{equation}}
\newcommand{\beqa}{\begin{eqnarray}}
\newcommand{\eeqa}{\end{eqnarray}}
\newcommand{\bea}{\begin{eqnarray}}
\newcommand{\eea}{\end{eqnarray}}
\newcommand{\dfrac}{\displaystyle \frac}
\renewcommand{\thefootnote}{\#\arabic{footnote}}
\newcommand{\ve}{\varepsilon}
\newcommand{\no}{\nonumber}
\newcommand{\krig}[1]{\stackrel{\circ}{#1}}
\newcommand{\barr}[1]{\not\mathrel #1}
\newcommand{\vs}{\vspace{-0.05cm}}
\def\w{\omega}

\begin{titlepage}
 
{\bf REVISED VERSION} \hfill {\tiny {\bf  FZJ-IKP(TH)-2000-06}}

\vspace{1.5cm}

\begin{center}

{\large  \bf {
Pion--nucleon scattering in chiral perturbation theory II:\\[0.2em]
Fourth order calculation\footnote{Work supported in part by a NATO
  research grant under no. BOT99/001.}}}

\vspace{1.2cm}
                              
{\large Nadia Fettes\footnote{email: N.Fettes@fz-juelich.de}
\footnote{Present address: Kellogg Radiation
  Laboratory, California Institute of Technology, 
 Pasadena, \hbox{CA 91125}, USA},
Ulf-G. Mei\ss ner\footnote{email: Ulf-G.Meissner@fz-juelich.de}
}

\vspace{1.0cm}

{\it Forschungszentrum J\"ulich, Institut f\"ur Kernphysik (Theorie)}

{\it D--52425 J\"ulich, Germany}\\

\end{center}

\vspace{0.8cm}

\begin{abstract}
\noindent We analyze elastic--pion nucleon scattering to fourth
order in heavy baryon chiral perturbation theory, extending the
third order study published in Nucl.~Phys. A640 (1998) 199. We use various
partial wave analyses to pin down the low--energy constants from
data in the {\em physical} region. The S--wave scattering lengths
are consistent with recent determinations from pionic hydrogen and 
deuterium. We find an improved description of the P--waves.
We also discuss the pion--nucleon sigma term and problems related
to the prediction of the  subthreshold parameters.

\end{abstract}

\vspace{2cm}


\vfill

\end{titlepage}

\section{Introduction and summary}
\label{sec:intro}
\def\theequation{\arabic{section}.\arabic{equation}}
\setcounter{equation}{0}

Effective field theory allows one to analyze the chiral structure of
Quantum Chromodynamics in the low energy domain, which is not accessible
to a perturbative expansion in the strong coupling constant. The spontaneous
violation of the chiral symmetry of QCD entails the existence of Goldstone
bosons, the pions (we consider here the two flavor case of the light
up and down quarks). The interactions of the Goldstone bosons among themselves
and with matter vanish as the momentum transfer goes to zero. This is a
consequence of Goldstone's theorem. Consequently, such interactions can be
analyzed in a perturbative expansion, where all momenta and masses are
small compared to the typical scale of hadronic interactions, say the mass
of the rho meson. This is the so--called chiral expansion.
A systematic investigation of processes involving
pions allows therefore to understand in a precise and quantitative
manner how the symmetry violation takes place and also to pin
down the ratios of the light quark masses. One such process is 
elastic pion--nucleon scattering.  In ref.\cite{fms} (called~(I) from here on),
we considered this reaction in the framework of heavy baryon chiral perturbation 
theory (HBCHPT) to third order in the chiral expansion. At that order, the
first contributions from pion loop graphs, which perturbatively restore
unitarity, appear. Besides pion loop diagrams, there are tree graphs. Some
of these have fixed coefficients, others are accompanied by coupling constants
not fixed by chiral symmetry. These so--called low energy constants (LECs)
must be determined by a comparison to data or can be estimated using models.
As has been shown in refs.\cite{EGPR,bkmlec}, these LECs encode the information
from higher mass states not present explicitly in the effective field theory.
There are three important reasons to extend the calculations of~(I) to fourth
order: First, only at that order one has a {\em complete} one--loop
representation. Second, it is known that these fourth order corrections can
be large (for a review see ref.\cite{bkmrev} and an update is given in
ref.\cite{ulflec}). Third, only after having obtained an accurate representation
of the isospin--symmetric amplitude, as done here, one can attack the more
subtle problem of isospin symmetry violation~\cite{gibbs,matsi}
in low energy pion--nucleon scattering.
First steps in the framework of HBCHPT have been reported in 
refs.\cite{ms,fmsi,mm}.

There have been some interesting new developments since~(I)
appeared: First, a manifestly Lorentz invariant form of baryon chiral
perturbation theory was proposed in ref.\cite{BL} and some implications
for the nucleon mass and the scalar form factor to fourth order
were worked out. The same group is also investigating pion--nucleon
scattering in that framework~\cite{BL2}. By construction, their approach 
leads to the correct analytical structure of the pion--nucleon scattering
amplitude, whereas in the heavy baryon approach special care has to be
taken in certain regions of the complex plane.
Second, a different determination of the dimension
two and three LECs by fitting the HBCHPT amplitude to the dispersive
representation (based on the Karlsruhe partial wave analysis) 
inside the Mandelstam triangle was performed in ref.\cite{paul}.
While that method has the a priori advantage that the chiral expansion
is expected to converge best in this special region of the Mandelstam plane,
it is difficult to work out the theoretical uncertainties. Another drawback 
of this procedure is that only
three dimension two LECs could be determined with sufficient precision
(if one uses the third order HBCHPT amplitude as input). This is related
to the fact that close to the point $\nu = t =0$ the contribution from
the third order terms is accompanied by very small kinematical prefactors.

Here, we will
follow the approach used in~(I), namely to fit to the phase shifts provided
by three different partial wave analyses for pion laboratory momenta between
40 and 100~MeV. This not only allows for a discussion of the uncertainties due 
to the input but also gives us the possibility to predict the phase shifts 
at higher and lower
momenta, in particular the  scattering lengths and the range parameters. 
Furthermore, we are then able  to directly compare to the third order calculation
and draw conclusions on the convergence of the chiral expansion. Of course,
at the appropriate places we will discuss the relation to the work reported 
in refs.\cite{BL,paul}.

\pagebreak

The pertinent results of the present investigation can be summarized as follows:
\begin{enumerate}
\item[(i)] We have constructed the complete one--loop amplitude for 
elastic pion--nucleon scattering in heavy baryon chiral
perturbation theory, including all terms of order $q^4$.  
It contains  13 low--energy constants plus one related to fixing
the pion--nucleon coupling constant through the Goldberger--Treiman
discrepancy. Their values can be
determined by fitting to the two S-- and four P--wave partial wave amplitudes
for three different sets of available pion--nucleon phase shifts in the 
physical region at low energies (typically in the range of 40 to 
100~MeV pion momentum in the laboratory frame). 
\smallskip
\item[(ii)]We have performed two types of fits. In the first one, we fit
four combinations of the dimension two and four LECs, together with five
LECs from the third and five from the fourth order. 
This means  that the dimension two LECs
are subject to quark mass renormalizations from certain fourth order
terms. Most fitted LECs are of ``natural'' size. In
the second approach, we fix the dimension two LECs as determined from the
third order fit and determine the corresponding dimension four LEC
combinations separately. We have studied the convergence of the chiral
expansion by comparing the best fits based on the second, third and fourth
order representation of the scattering
amplitudes. The fourth order corrections are in general not large, but they
improve the description of most partial waves. This indicates convergence
of the chiral expansion.
\smallskip
\item[(iii)]We can  predict the phases at {\it lower} and at 
{\it higher} energies, in particular the threshold parameters 
(scattering lengths and effective ranges). The results are not very
different from the third order study in~(I), but the description of the
P--waves is improved, in particular  the scattering length in the
delta channel and the energy dependence of the small P--waves. 
The errors  on the S-wave scattering lengths are as in~(I) since
they are due to  the  differences in the partial wave analyses used
as input. Our theoretical predictions are consistent with recent determinations
from pionic hydrogen and deuterium~\cite{PSI}.
\smallskip
\item[(iv)] The pion--nucleon sigma term (at zero momentum
transfer) can not be predicted without further
input since at fourth order a new combination of LECs appears, that is not
pinned down by the  scattering data. Therefore, we have analyzed
the sigma term at  the Cheng--Dashen point. Using a family of
sum rules which relate this quantity to threshold parameters and known
dispersive integrals,
we find results consistent with other determinations using the various
partial wave analyses. 
\smallskip
\item[(v)] We do not find any improvement of the chiral description of
the so--called  subthreshold parameters as reported in~(I). In some cases, the
fourth order prediction is worse than the third order one. Since our 
amplitude is pinned down in the physical region, we do not expect the
extrapolation to the subthreshold region to be very precise.
To circumvent this problem, it is
mandatory to combine the chiral representation obtained here with dispersion
relations, see e.g. ref.\cite{BL2}, or by fitting directly inside the
Mandelstam triangle~\cite{paul}.
\end{enumerate} 

\medskip

The manuscript is organized as follows. In section~2, we briefly discuss
the effective Lagrangian underlying our calculation. All details were
given in~(I), so here we only spell out the new terms at fourth order.
Section~3 contains the HBCHPT results for the pion--nucleon scattering 
amplitudes $g^\pm, h^\pm$ to fourth order. The fitting procedure together
with the results for the phase shifts and threshold parameters are 
presented in section~4. We also spell out the remaining problems related
to the sigma term and the subthreshold parameters. The appendix contains
the analytical expressions for the various threshold parameters.

\section{Effective Lagrangian}
\setcounter{equation}{0}

The starting point of our approach is the most general
chiral invariant Lagrangian built from pions, nucleons and external
scalar sources (to account for the explicit chiral symmetry breaking). 
The Goldstone bosons are collected in a 2$\times$2 matrix-valued field
$U(x)= u^2(x)$. We use the so--called sigma model parametrisation (gauge).
We work in the framework of heavy baryon chiral perturbation theory, thus
the nucleons are described by structureless non--relativistic spin-${\small
  \frac{1}{2}}$ particles, denoted by $N(x)$.
The effective theory admits a low energy expansion, i.e. the
corresponding effective Lagrangian can be written as (for more details
and references, see e.g. \cite{bkmrev})
\beq
{\cal L}_{\rm eff} = 
 {\cal L}_{\pi\pi}^{(2)} +  {\cal L}_{\pi\pi}^{(4)} +
 {\cal L}_{\pi N}^{(1)}  +  {\cal L}_{\pi N}^{(2)} +
 {\cal L}_{\pi N}^{(3)}  +  {\cal L}_{\pi N}^{(4)}  \ldots~,
\eeq
where the ellipsis denotes terms of higher order. For the explicit
form of the meson Lagrangian and the dimension one, two and three
pion--nucleon terms, we refer to~(I). The complete fourth order
Lagrangian is given in ref.\cite{FMSM} and the renormalization is
discussed in ref.\cite{MMS}. For completeness, we display here the
finite terms from ${\cal L}_{\pi N}^{(4)}$ which contribute to elastic $\pi$N
scattering
\begin{eqnarray}
{{\cal L}}_{\pi N}^{(4)} &=&\bar{N} \Bigg\{ 
\Big(\bar{e}_{14} -\frac{1}{16 m^2} c_2 \Big)
    \left\langle h_{\mu\nu }h^{\mu \nu }\right\rangle 
+\Big(\bar{e}_{15} -\frac{1}{256 m^3}g_{A}^{2}-\frac{1}{16 m}\left( \bar{d}_{14}-\bar{d}_{15}\right)\Big)
    v^{\mu }v^{\nu }\left\langle h_{\lambda \mu}h_{\;\;\nu }^{\lambda }\right\rangle 
\no\\&&
+\Big(\bar{e}_{16} + \frac{3}{256 m^3}g_{A}^{2}\Big) 
   v^{\lambda }v^{\mu }v^{\nu }v^{\rho}\left\langle h_{\lambda \mu }h_{\nu \rho }\right\rangle 
+\bar{e}_{17}[S^{\mu },S^{\nu }]\left[ h_{\lambda \mu },h_{\;\;\nu}^{\lambda }\right] 
\no\\&&
+\Big(\bar{e}_{18}-\frac{1}{128  m^3}\left(2+g_{A}^{2}\right) -\frac{1}{16 m^2} c_{4} \Big)
   [S^{\mu },S^{\nu }]v^{\lambda }v^{\rho }\left[h_{\lambda \mu },h_{\nu \rho }\right] 
+\bar{e}_{19} \left\langle \chi _{+}\right\rangle \left\langle u\cdot u\right\rangle 
\no\\&&
+\Big(\bar{e}_{20} -\frac{1}{32 m^2}g_{A}^{2}c_{1}-\frac{1}{8 m} g_{A} \bar{d}_{16}\Big)
   \left\langle \chi_{+}\right\rangle \left\langle (v\cdot u)^{2}\right\rangle 
+\Big(\bar{e}_{21} + \frac{1}{16 m^2}c_{1}\Big)
   [S^{\mu },S^{\nu }]\left\langle \chi_{+}\right\rangle \left[ u_{\mu },u_{\nu }\right] 
\no\\&&
+\bar{e}_{22} \left[ D_{\mu },\left[ D^{\mu },\left\langle \chi_{+}\right\rangle \right] \right] 
+\bar{e}_{35} i v^{\mu }v^{\nu }\left\langle \widetilde{\chi }_{-}h_{\mu \nu }\right\rangle 
+\bar{e}_{36} i\left\langle u_{\mu }\left[ D^{\mu },\widetilde{\chi }_{-}\right] \right\rangle 
\no\\&&
+\bar{e}_{37} i[S^{\mu },S^{\nu }]\left[ u_{\mu },\left[ D_{\nu },\widetilde{\chi }_{-}\right] \right] 
+\bar{e}_{38} \left\langle \chi _{+}\right\rangle \left\langle \chi_{+}\right\rangle 
+\bar{e}_{115} \langle \chi \chi^{\dagger }\rangle
+\bar{e}_{116} \big(\det \chi +\det \chi ^{\dagger }\big)
\no\\&&
-\frac{1}{16 m^2}  c_2  \left\langle h_\mu^{\;\mu} h_\nu^{\;\nu} \right\rangle 
-\Big(\frac{1}{128 m^3}g_{A}^{2}-\frac{1}{16 m}\left(\bar{d}_{14}-\bar{d}_{15}\right) \Big)
   v^{\mu }v^{\nu}\left\langle h_{\mu \nu }h_{\rho }^{\;\rho }\right\rangle 
\no\\&&
-\frac{1}{4 m^2} c_2 \left\langle u_\mu [D^\mu,h_\nu^{\;\nu}]\right\rangle
-\Big(\frac{1}{128 m^3}g_{A}^{2}+\frac{1}{8m}\left(\bar{d}_{14}-\bar{d}_{15}\right)\Big)
   \left\langle v\cdot u\left[v\cdot D,h_{\mu }^{\;\mu }\right] \right\rangle 
\no\\&&
-\Big(\frac{1}{128 m^3} g_{A}^{2}-\frac{1}{8 m}\left( \bar{d}_{14}-\bar{d}_{15}\right) \Big)
   v^{\mu }\langle u^{\nu }\left[ v\cdot D,h_{\mu \nu }\right] \rangle 
+\frac{1}{32 m^3}g_{A}^{2}
   v^{\mu }v^{\nu}\left\langle v\cdot u\left[ v\cdot D,h_{\mu \nu }\right] \right\rangle 
\no\\&&
+\Big(\frac{1}{32 m^3}\left( 1+g_{A}^{2}\right) +\frac{1}{8 m^2} c_{4}\Big)
[S^{\mu },S^{\nu }]v^{\rho}\left[ u_{\nu },\left[ v\cdot D,h_{\mu \rho }\right] \right] 
-\frac{1}{8 m}g_{A}\bar{d}_{18}
i\left\langle v\cdot u\left[ v\cdot D,\widetilde{\chi }_{-}\right] \right\rangle \no
\no\\&&
+\Big(\frac{1}{128}g_{A}^{2}+\frac{1}{32 m^2}c_{4}\Big)
\Big( \left[ h_{\mu }^{\;\mu},u^{\lambda }\right] D_\lambda +{\rm h.c.} \Big)
\no\\&&
-\Big(\frac{1}{128 m^3}g_{A}^{2}+\frac{1}{32 m^2} c_{4}-\frac{1}{4 m}\left( \bar{d}_{1}+\bar{d}_{2}\right)\Big)
\Big([h^{\lambda\mu},u_{\mu}]D_\lambda +{\rm h.c.} \Big)
\no\\&&
+\Big(\frac{1}{128 m^3}\left( g_{A}^{2}-1\right)+\frac{1}{2 m}\bar{d}_{3}\Big)
\Big(v_{\mu }[h^{\lambda \mu},v\cdot u]D_\lambda +{\rm h.c.} \Big)
\no\\&&
+\Big(\frac{1}{128 m^3}+\frac{1}{4m}\bar{d}_{3}\Big)
\Big(v^{\mu }v^{\nu }[h_{\mu\nu },u^{\lambda }]D_\lambda +{\rm h.c.} \Big)
\no\\&&
+\Big(\frac{1}{64 m^3}g_{A}^{2}-\frac{1}{8 m^2}c_{3}\Big)
\Big([S^{\lambda },S^{\mu}]\langle h_{\mu \nu }u^{\nu }\rangle D_\lambda +{\rm h.c.} \Big)
\no\\&&
+\frac{1}{4 m}\left( \bar{d}_{14}-\bar{d}_{15}\right) 
\Big([S^{\mu },S^{\nu}]\langle h^{\lambda}_{\,\, \mu }u_{\nu }\rangle D_\lambda +{\rm h.c.} \Big)
\no\\&&
-\Big(\frac{1}{64 m^3}g_{A}^{2}+\frac{1}{8 m^2}c_{2}-\frac{1}{4 m}\left( \bar{d}_{14}-\bar{d}_{15}\right) \Big)
\Big([S^{\lambda },S^{\mu}]v^{\nu }\langle h_{\mu \nu }v\cdot u\rangle D_\lambda +{\rm h.c.} \Big)
\no\\&&
+\Big(\frac{1}{64 m^3}g_{A}^{2}-\frac{1}{4 m}\left( \bar{d}_{14}-\bar{d}_{15}\right)\Big) 
\Big([S^{\lambda },S^{\mu}]v^{\nu }v^{\rho }\left\langle u_{\mu }h_{\nu \rho }\right\rangle 
D_\lambda +{\rm h.c.} \Big)
\no\\&&
+\frac{1}{32 m^3}g_{A}
\Big(iS^{\lambda }\left[v\cdot D,h_{\mu }^{\;\mu }\right] D_\lambda +{\rm h.c.} \Big)
+\frac{1}{16 m^3}g_{A}
\Big(i\left[ S\cdot D,h_{\mu}^{\;\mu }\right] v\cdot D +{\rm h.c.} \Big)
\no\\&&
-\frac{1}{32 m^3} g_{A}
\Big(iS^{\nu }\left[ D^{\mu},h_{\mu \nu }\right] v\cdot D +{\rm h.c.} \Big)
-\frac{1}{32 m^3}g_{A}
\Big(iS^{\mu }v^{\nu }\left[v\cdot D,h_{\mu \nu }\right] v\cdot D +{\rm h.c.} \Big)
\no\\&&
-\frac{1}{32 m^3}g_{A}
\Big(iS^{\lambda }v^{\mu}v^{\nu }\left[ v\cdot D,h_{\mu \nu }\right] D_\lambda +{\rm h.c.} \Big)
\no\\&&
-\Big(\frac{1}{4 m^2} g_{A}c_{1}+\frac{1}{2 m}\bar{d}_{16}\Big)
\Big(iS^{\lambda }\langle \chi_{+}\rangle v\cdot u D_\lambda +{\rm h.c.} \Big)
-\frac{1}{4 m^2}c_{1}
\Big([S^{\lambda },S^{\mu }]\left[ D_{\mu },\left\langle \chi _{+}\right\rangle \right] 
D_\lambda +{\rm h.c.} \Big)
\no\\&&
+\frac{1}{2m}\bar{d}_{5}
\Big(i[\widetilde{\chi }_{-},u^{\lambda }]D_\lambda +{\rm h.c.} \Big)
+\frac{1}{2m}\bar{d}_{18}
\Big(S^{\lambda }[v\cdot D,\widetilde{\chi }_{-}]D_\lambda +{\rm h.c.} \Big)
\no \\
&&
-\frac{1}{2m^2} c_2
D_\mu \left\langle u^\mu u^\nu \right\rangle D_\nu
+\Big(\frac{1}{64 m^3}g_{A}^{2}-\frac{1}{8 m^2}c_{3}\Big)
D_\mu \left\langle u\cdot u\right\rangle D^\mu
\no\\&&
-\Big(\frac{1}{64 m^3}g_{A}^{2}+\frac{1}{8 m^2} c_{2}\Big)
D_\mu \left\langle\left( v\cdot u\right) ^{2}\right\rangle D^\mu
-\Big(\frac{1}{32 m^3}g_{A}^{2}+\frac{1}{8 m^2} c_{4}\Big)
D_\mu \left[ S^{\rho},S^{\tau }\right] \left[ u_{\rho },u_{\tau }\right] D^\mu
\no\\&&
+\Big(\frac{1}{16 m^3}g_{A}^{2}+\frac{1}{4 m^2}c_{4}\Big) 
\Big(D_\mu [S^{\mu },S^{\rho }]\left[u^{\nu },u_{\rho }\right] D_\nu + {\rm h.c.} \Big)
-\frac{1}{4 m^2}c_{1} 
D_\mu \left\langle\chi _{+}\right\rangle D^\mu
\no\\&&
-\frac{1}{4 m^3}g_{A}
\Big( iu\cdot D S\cdot D v\cdot D +{\rm h.c.} \Big)
+\frac{1}{8 m^3}g_{A}
\Big( iS\cdot u D^2 v\cdot D +{\rm h.c.} \Big)
\\&&
+\frac{3}{8 m^3}g_{A}
\Big( iv\cdot u S\cdot D v\cdot D v\cdot D +{\rm h.c.} \Big)
-\frac{1}{8 m^3}g_{A}
\Big( iS\cdot u v\cdot D v\cdot D v\cdot D +{\rm h.c.} \Big)
\Bigg\}{N}~, \no
\end{eqnarray}
with $v_\mu$ the nucleon's four--velocity, $S_\mu$ the covariant
spin--vector, $D_\mu$ the chiral covariant derivative,
$u_\mu = i(\partial_\mu u u^\dagger + u^\dagger \partial_\mu u)$, $h_{\mu\nu} =
[D_\mu,u_\nu]+[D_\nu,u_\mu]$ and $\chi_\pm = u^\dagger \chi u^\dagger
\pm u \chi^\dagger u$ encoding the explicit chiral symmetry breaking.
Traces in flavor space are denoted by $\langle \ldots \rangle$ and
$\tilde{\chi}_- = \chi_- - \langle \chi_- \rangle/2$. 
We remark that the various parameters like $g_A, m, \ldots$ 
appearing in the effective
Lagrangian have to be taken in the chiral SU(2) limit ($m_u=m_d=0\, ,
\,\, m_s$ fixed) and should be denoted as $\krig{g}_A , \krig{m},
\ldots$. Throughout this manuscript, we will not specify this but it
should be kept in mind.  It is also worth mentioning the particular
role of the terms $\sim e_{115,116}$. These operators have no pion
matrix elements but are simply contact interactions of the external
scalar source with the matter fields and thus contribute to the
nucleon mass and the scalar form factor. This will be of importance
later on. 
Having constructed the effective pion--nucleon Lagrangian to order $q^4$, 
we now turn to use it in order to describe elastic pion--nucleon scattering. To account 
for isospin breaking, one  has to extend this Lagrangian to include
virtual photons. This has already been done in~\cite{ms,mm} and we refer the
reader to these papers. For a systematic study of isospin violation in the
elastic and charge exchange channels, one first has to find out to what 
accuracy the low energy $\pi N$ phase shifts can be described in the
isospin symmetric framework.\footnote{An investigation of isospin
violation in the threshold
amplitudes to third order was reported in ref.\cite{fmsi}.}
This is the question which will be addressed in
the remaining sections of this investigation.

\section{Pion--nucleon scattering}
\setcounter{equation}{0}

\subsection{Basic definitions}

In this section, we only give a few basic definitions pertinent to
elastic pion--nucleon scattering. For a more detailed discussion, we
refer to~(I). In the center-of-mass system (cms), the amplitude for the process
$\pi^a(q_1) + N(p_1) \to \pi^b(q_2) + N(p_2)$ takes the
following form (in the isospin basis): 
\beqa 
T^{ba}_{\pi N} &=& 
\biggl(\frac{E+m}{2m}\biggr) \, \biggl\lbrace \delta^{ba} 
\Big[ g^+(\omega,t)+ i \vec
\sigma \cdot(\vec{q}_2\times \vec{q}_1\,) \, h^+(\omega,t) \Big]
\nonumber \\ && \qquad\quad
+i \, \epsilon^{bac}
\tau^c \Big[ g^-(\omega,t)+ i \vec \sigma \cdot(\vec{q}_2 \times \vec{q}_1\,) \,
h^-(\omega,t) \Big] \biggr\rbrace 
\eeqa
with $\omega = v\cdot q_1 = v\cdot q_2$ the pion cms energy, 
$E_1 = E_2 \equiv E = ( \vec{q\,}^2 +m^2)^{1/2}$ the nucleon energy and
$\vec{q\,}_1^2 = \vec{q\,}_2^2 \equiv \vec{q\,}^2 = ((s-M^2-m^2)^2 -4m^2M^2)/ (4s)$.
$t=(q_1-q_2)^2$ is the invariant momentum transfer squared and $s$ 
denotes the total cms energy squared. 
Furthermore, $g^\pm(\omega,t)$ refers to the
isoscalar/isovector non-spin-flip amplitude and $h^\pm(\omega,t)$ to the
isoscalar/isovector spin-flip amplitude. This form is most suitable
for the HBCHPT calculation since it is already defined in a
two--component framework.

\subsection{Chiral expansion of the amplitudes}

What we are after is the chiral expansion of the various amplitudes
$g^\pm, h^\pm$. These consist of essentially three pieces, which are
the tree and counterterm parts of polynomial type as well as the
unitarity corrections due to the pion loops. To be precise, we
have
\beq\label{chexp}
X = X^{\rm tree} + X^{\rm ct} + X^{\rm loop} \,\, , \quad
X= g^\pm  ,  h^\pm \,\, ,
\eeq
where the tree contribution subsumes all Born terms with fixed coefficients, 
the counterterm amplitude the ones
proportional to the dimension two, three and four LECs.
The last term in Eq.(\ref{chexp}) is the complete one--loop
amplitude consisting of terms of order $q^3$ and $q^4$. 
The latter is a complex--valued function and
restores unitarity in the perturbative sense. Its various terms
are all proportional to $1/F^4$. We remark that the topologies of the
new loop graphs are not different from the ones already present at third
order. The loops of fourth order have exactly one insertion from the
dimension two Lagrangian.
Note that we treat the chiral symmetry breaking scale $\Lambda_\chi \simeq
1\,$GeV on the same footing as the nucleon mass. In principle, one
could also organize the loop expansion, which proceeds in powers of
$1/F^2$, and the $1/m$ expansion independently of each other (with some
prescription for the mixed terms). 
These amplitudes are functions of two kinematical variables,
which we choose to be the pion energy and the invariant
momentum transfer squared, i.e. $X= X (\omega, t)$. In what
follows, we mostly suppress these arguments.
The full one--loop amplitude to order $q^4$ is obtained after mass
and coupling constant renormalization,
\beq
(\krig{g}_A , \krig{m}, F , M) \to ({g}_A , {m}, F_\pi , M_\pi )~.
\eeq
To fourth order, these read (we also give the corresponding $Z$--factors
for the pion and the nucleon)
\bea
Z_{\pi} & = & 1-\frac{2 M^2}{F^2} {\ell}_4 - \frac{\Delta_\pi}{F^2}~, \\
Z_N &=& 1- \frac{g_A^2}{F^2} \frac{3 M^2}{32 \pi^2} 
+ \frac{g_A^2}{F^2} \frac{9 M^3}{64 \pi m}~,\\
M^2_\pi & = & M^2 \left\{ 1 +\frac{2 M^2}{F^2} {\ell}_3 +\frac{\Delta_\pi}{2 F^2} 
\right\}~,\\
m &=& \krig{m} -4 M^2 c_1 -\frac{3 g_A^2 M^3}{32 \pi F^2} \no\\
&& \qquad -  M^4 (16\bar{e}_{38} + 2\bar{e}_{115}+ \frac{1}{2}\bar{e}_{116}) 
+ \frac{3 M^4 c_2}{128 \pi^2 F^2} - \frac{3 g_A^2 M^4}{64 \pi^2 m
  F^2}~,\\
F_\pi & = & F \left\{ 1 + \frac{M^2}{F^2} \ell_4 -\frac{\Delta_\pi}{F^2}\right\}~, \\
\frac{g_A}{F_{\pi}} & = & \frac{\krig{g}_A}{F} \Bigg\{ 1 -\frac{M^2}{F^2} \ell_4 +
\frac{4 M^2}{g_A} d_{16}(\lambda)  + \frac{g_A^2}{4 F^2}
\left( \Delta_\pi -\frac{M^2}{4 \pi^2} \right)  \no\\&& \hspace{2cm}
-\frac{M^3}{6 \pi F^2} \left( -\frac{1}{8m} + c_3 - \left( 2 c_4 +\frac{1}{2m} \right)
-\frac{3 g_A^2}{4m} \right)
\Bigg\}~,
\eea
with 
\beq
\Delta_\pi = 2 M_\pi^2 \left( L +\frac{1}{16 \pi^2} \ln{\frac{M_\pi}{\lambda}} \right) + {\cal O}(d-4)
\eeq
and 
\beq
L=\frac{\lambda^{d-4}}{16 \pi^2} \left( \frac{1}{d-4} +\frac{1}{2} (\gamma_E -1 -\ln{ 4\pi}) \right) ~,
\eeq
where Euler's constant $\gamma_E = 0.557215$ has been used,
$\lambda$ is the scale of dimensional regularization and $d$
the number of space--time dimensions.

\medskip\noindent
After these preliminaries, we give the final expressions for the tree,
counterterm and loop graphs at fourth order in terms of the renormalized quantities:

\smallskip\noindent
\underline{Tree and counterterm graphs:}

\bea
m^3 F_\pi^2 g^+(\w,t) & = & -\frac{g_A^2}{64\omega^4} \Big[ -32 \omega^4 M_\pi^2 t+45 M_\pi^4 t^2 
-11 M_\pi^2 t^3 -80 M_\pi^6 t +52 M_\pi^8
\no\\&&\hspace{1.2cm}+7 \omega^2 t^3 +110 \omega^2 M_\pi^4 t-49 \omega^2 M_\pi^2 t^2 -76 \omega^2 M_\pi^6+11 \omega^4 t^2
\no\\&&\hspace{1.2cm}+t^4-4 \omega^6 t+28 \omega^4 M^4 \Big]
\no\\&&+\frac{1}{2} M_\pi^2 (2 \w^2 - 2 M_\pi^2 + t) m  c_1 
+8 M_\pi^4 m^3 c_1 \frac{\ell_3}{F_\pi^2}\no\\&&
+\frac{1}{4}(-22 \w^2 M_\pi^2 + 8 M_\pi^4 + 3 \w^2 t +14 \w^4 -4 M_\pi^2 t) m c_2\no\\&&
+\frac{1}{8} (-4 \w^2 M_\pi^2 +2 \w^2 t +4 M_\pi^4 -4 M_\pi^2 t + t^2) m c_3
-16 M_\pi^2 \w^2 m^2 c_1 c_2\no\\&&
+ \w^2 t m^2 (\bar{d}_{14}-\bar{d}_{15})
+\frac{g_A}{4 \w^2} M_\pi^2 (4 M_\pi^4 + t^2 + 4 \w^2 t -4 M_\pi^2 t) m^2 \bar{d}_{18}\no\\&&
-4 (4 M_\pi^2 t -4 M_\pi^4 +t^2) m^3 \bar{e}_{14}
-8 \w^2 (-2 M_\pi^2 +t) m^3 \bar{e}_{15}\no\\&&
+16 \w^4 m^3 \bar{e}_{16}
-4 M_\pi^2 (-2 M_\pi^2 +t) m^3 (2 \bar{e}_{19}-\bar{e}_{22}-\bar{e}_{36})\no\\&&
+16 \w^2 M_\pi^2 m^3 (\bar{e}_{20} + \bar{e}_{35} -\frac{g_A \bar{d}_{16}}{8 m} )
+8 M_\pi^4 m^3 (\bar{e}_{22}-4 \bar{e}_{38})~,\\
m^3 F_\pi^2 h^+(\w,t) & = & 
\frac{g_A^2}{32\omega^4} \Big[ 5 \omega^2 t^2 +27 M_\pi^4 t 
-9 M_\pi^2 t^2 -25 \omega^2 M_\pi^2 t - 28 M_\pi^6 +t^3 \no\\&&\hspace{1.2cm}
+30 \omega^2 M_\pi^4 -4 \omega^4 M_\pi^2 +3 \omega^4 t \Big]\no\\&&
+M_\pi^2 m c_1 
-\frac{1}{2} \w^2 m c_2 
+\frac{1}{4} (-2 M_\pi^2 +t) m c_3\no\\&&
+\frac{1}{2} (8\w^2-4 M_\pi^2 +t) m^2 (\bar{d}_{14}-\bar{d}_{15})
-\frac{g_A}{2 \w^2} M_\pi^2 (-4 M_\pi^2+t) m^2 \bar{d}_{18}~,\\
m^3 F_\pi^2 g^-(\w,t) & = & 
\frac{g_A^2}{64 \omega^4} \Big[ -32 \omega^4 M_\pi^2 t + 45 M_\pi^4 t^2 
-11 M_\pi^2 t^3 -82 M_\pi^6 t +56 M_\pi^8 \no\\&&\hspace{1.2cm}
+7 \omega^2 t^3 +110 \omega^2 M_\pi^4 t-49 \omega^2 M_\pi^2 t^2 -80 \omega^2 M_\pi^6 +11 \omega^4 t^2 
\no\\&&\hspace{1.2cm}
+ 8 \omega^8+ t^4 -2 \omega^6 t -8 \omega^6 M_\pi^2 +24 \omega^4 M_\pi^4 \Big]\no\\&&
+\frac{1}{32\omega^4} \Big[-4 \omega^8+8 \omega^6 M_\pi^2 -4 \omega^4 M_\pi^4 
- \omega^6 t + \omega^4 M_\pi^2 t \Big]\no\\&&
+\frac{1}{16} t (4 \w^2 - 4 M_\pi^2 +t) m c_4\no\\&&
-\frac{1}{2} (-8\w^2 M_\pi^2 +4 \w^2 t +8 M_\pi^4 +t^2 -6 M_\pi^2 t) m^2 (\bar{d}_1+\bar{d}_2)\no\\&&
+3 \w^2 (4\w^2-4 M_\pi^2 +t) m^2 \bar{d}_3
+2 M_\pi^2 (4\w^2 -4 M_\pi^2 +t) m^2 \bar{d}_5\no\\&&
-\frac{g_A}{4\w^2} M_\pi^2 (t^2 +2 \w^2 t -8 \w^4 +8 M_\pi^4 -6 M_\pi^2 t) m^2 \bar{d}_{18}~,\\
m^3 F_\pi^2 h^-(\w,t) & = & 
\frac{-g_A^2}{32\omega^4} \Big[ 5 \omega^2 t^2 + 27 M_\pi^4 t 
-9 M_\pi^2 t^2 -25 \omega^2 M_\pi^2 t-26 M_\pi^6 +t^3\no\\&&\hspace{1.2cm}
+30 \omega^2 M_\pi^4 -4 \omega^4 M_\pi^2+3 \omega^4 t -2 \omega^6 \Big] 
-\frac{\omega^4(\omega^2-M_\pi^2) }{16 \omega^4} \no\\&&
- M_\pi^2 m c_1
+\frac{1}{8} (2 \w^2 -2 M_\pi^2 +t) m c_4
+\frac{g_A}{2 \w^2} M_\pi^2 (2 \w^2-2 M_\pi^2 +t) m^2 \bar{d}_{18}\no\\&&
-4 (-2 M_\pi^2 +t) m^3 \bar{e}_{17}
+8 \w^2 m^3 \bar{e}_{18}
+4 M_\pi^2 m^3 (2 \bar{e}_{21} -\bar{e}_{37})~.
\eea

\smallskip\noindent
\underline{Loop graphs:}
\bea
m F_\pi^4 g^+(\w,t) & = & 
-\frac{1}{12 \w^3} J_0(\w) \Big[2 M_\pi^2 g_A^4 (t M_\pi^2 -2 \w^4 +4 \w^2 M_\pi^2 -t \w^2 -2 M_\pi^4)
      \no\\&&\hspace{2.2cm}+\w^2 g_A^2 (-12 \w^2 M_\pi^2 - M_\pi^2 t +\w^2 t +12 \w^4)+6 \w^4 M_\pi^2
      \Big]\no\\&&
+\frac{1}{12 \w^3} J_0(-\w) \Big[2 g_A^4 (6 M_\pi^6 +8 \w^6-10 \w^4 M_\pi^2 +3 t \w^2 M_\pi^2 
      + M_\pi^2 t^2 +2 t \w^4 
      \no\\&&\hspace{3cm}-4 \w^2 M_\pi^4 -5 t M_\pi^4)-\w^2 g_A^2 (M_\pi^2 t +4 \w^2 M_\pi^2 +\w^2 t -4 \w^4) 
      \no\\&&\hspace{2.4cm}
      +6 \w^4 (-3 M_\pi^2 +t +4 \w^2)\Big]\no\\&&
-\frac{1}{12\w^2} \frac{\partial J_0}{\partial \w} (\w) 
     \Big[ g_A^4 (8 \w^6 + \w^2 t^2 - 8 M_\pi^6 - M_\pi^2 t^2 +24 \w^2 M_\pi^4 +6 t \w^4 +6 t M_\pi^4 
     \no\\&&\hspace{3cm}-12 t \w^2 M_\pi^2 -24 \w^4 M_\pi^2 )
     \no\\&&\hspace{2.2cm}+2 \w^2 g_A^2 ( 4 \w^4 - M_\pi^2 t -8 \w^2 M_\pi^2 +\w^2 t +4 M_\pi^4)
     \no\\&&\hspace{2.2cm}+3 \w^4 (4 \w^2 -4 M_\pi^2 +t)\Big]\no\\&&
+\frac{g_A^2}{32} \frac{\partial K_0}{\partial \w}(0,t) (12 M_\pi^4 t -4 M_\pi^6 -9 M_\pi^2 t^2 +2 t^3)
     \no\\&&
-\frac{1}{24} I_0(t) \Big[ 3 g_A^2 t (2 t-M_\pi^2)
     -48 M_\pi^2 m c_1 (M_\pi^2 -2 t)
     \no\\&&\hspace{1.8cm}+2 m c_2 (2 t^2 +4 M_\pi^4 - 9 t M_\pi^2)
     +12 m c_3 (-5 t M_\pi^2 +2 t^2 + 2 M_\pi^4)\Big]\no\\&&
-\frac{1}{1152 \pi^2 \w^3} \Big[ 2 g_A^4 ( -52 \w^5 M_\pi^2 +24 \w^7+\w^3 t^2+10 t \w^5 
     -96 \pi M_\pi^7 +96 \pi M_\pi^5 \w^2
     \no\\&&\hspace{2.8cm}+72 \pi M_\pi^5 t -12 \pi M_\pi^3 t^2 -48 \pi M_\pi^3 \w^2 t 
     -12 \w M_\pi^4 t +6 \w M_\pi^2 t^2 
     \no\\&&\hspace{2.8cm}+11 \w^3 M_\pi^2 t +46 M_\pi^4 \w^3)\no\\&&\hspace{2.0cm}
     +g_A^2 ( -18 \w^3 t^2 -288 \w^5 M_\pi^2 +60 \w^3 M_\pi^4 +33 \w^3 t M_\pi^2 +192 \w^7 
     \no\\&&\hspace{2.8cm}+64 \w^5 t )
     +4 \w^3 m c_2 (2 t^2 +6 M_\pi^4 -13 M_\pi^2 t)\Big]~,\\
m F_\pi^4 h^+(\w,t) & = & 
\frac{g_A^2}{12 \w^3} J_0(\w) (\w^2-M_\pi^2) \Big[g_A^2 M_\pi^2 -2 \w^2 +8 m \w^2 (c_3-c_4) \Big]
     \no\\&&
-\frac{g_A^2}{12 \w^3} J_0(-w) \Big[g_A^2 (M_\pi^2 \w^2 + M_\pi^2 t -3 M_\pi^4 +2 \w^4) 
     \no\\&&\hspace{2.8cm}+(\w^2-M_\pi^2)(2\w^2 -8 \w^2 m (c_3-c_4))\Big]\no\\&&
+\frac{g_A^4}{24 \w^2} \frac{\partial J_0}{\partial \w}(-\w)    
     (4 \w^4 -M_\pi^2 t -8 \w^2 M_\pi^2 +\w^2 t +4 M_\pi^4)\no\\&&
-\frac{g_A^2}{32} \frac{\partial K_0}{\partial \w} (0,t) (-9 M_\pi^2 t +4 M_\pi^4 +2 t^2)
-\frac{g_A^2}{8} I_0(t) (2 t -M_\pi^2)\no\\&&
-\frac{g_A^2}{1152 \w^3 \pi^2} \Big[2 g_A^2(-12 \pi M_\pi^5 -6 \pi M_\pi^3 \w^2 +6 \pi M_\pi^3 t 
     +32 \w^5 +4 \w^3 t 
     \no\\&&\hspace{2.8cm}-16 \w^3 M_\pi^2 +12 \w M_\pi^4 -3 \w t M_\pi^2)
     \no\\&&\hspace{2.0cm}+3 \w^2 (3 \w M_\pi^2 -6 \w t -16 \pi M_\pi^3) 
     +192 \pi M_\pi^3 \w^2m (c_3-c_4)\Big]~,\\
m F_\pi^4 g^-(\w,t) & = &  
\frac{1}{24 \w^3} J_0(\w) \Big[ g_A^4 (t \w^2 M_\pi^2 +2 \w^4 M_\pi^2 -4 \w^2 M_\pi^4 -t M_\pi^4 +2 M_\pi^6)\
     \no\\&&\hspace{1.8cm}+g_A^2 (-6 \w^2 M_\pi^4 +12 \w^4 M_\pi^2 +3 \w^2 M_\pi^2 t 
     -3 \w^4 t -12 \w^6
     \no\\&&\hspace{2.6cm}+8 \w^2 m (c_3-c_4) (2 \w^4-4 M_\pi^2 \w^2+2 M_\pi^4 - t M_\pi^2 +\w^2 t))
     \no\\&&\hspace{1.8cm}+6 \w^4 (-M_\pi^2-16 M_\pi^2 m c_1+8 \w^2 m (c_2 +c_3))\Big]\no\\&&
-\frac{1}{24 \w^3} J_0(-\w)\Big[g_A^4 (-5 t M_\pi^4 -10 \w^4 M_\pi^2 -4 \w^2 M_\pi^4 
     +3 t \w^2 M_\pi^2 +6 M_\pi^6
     \no\\&&\hspace{3.1cm}+8 \w^6+M_\pi^2 t^2 +2 t \w^4)
     \no\\&&\hspace{2.3cm}+g_A^2 \w^2 (6 M_\pi^4 -4 \w^2 M_\pi^2 +\w^2 t -3 M_\pi^2 t +4 \w^4 
     \no\\&&\hspace{3.6cm}+8 m (c_3-c_4) (-2 M_\pi^4 +t M_\pi^2 -\w^2 t -2 \w^4 +4 \w^2 M_\pi^2))
     \no\\&&\hspace{2.3cm}+6 \w^4 (-3 M_\pi^2 +t +4 \w^2 + 16 M_\pi^2 m c_1 - 8 \w^2 m (c_2 +c_3)
     )\Big]\no\\&&
+\frac{1}{48 \w^2} \frac{\partial J_0}{\partial\w}(-\w) 
    \Big[ g_A^4 (-12 t \w^2 M_\pi^2 -24 \w^4 M_\pi^2 +6 t M_\pi^4 +8 \w^6+24 \w^2 M_\pi^4 
    \no\\&&\hspace{3.6cm}-8 M_\pi^6+6 t \w^4 -M_\pi^2 t^2 +\w^2 t^2)
    \no\\&&\hspace{3.1cm}+4 \w^2 g_A^2 (4 \w^4 +4 M_\pi^4 +\w^2 t -8 \w^2 M_\pi^2 - M_\pi^2 t)
    \no\\&&\hspace{3.1cm}+6 \w^4 (-4 M_\pi^2 +4 \w^2 +t)\Big]\no\\&&
+\frac{g_A^2}{8} K_0(0,t) \w (4 M_\pi^4 +t^2 -4 M_\pi^2 t)\no\\&&
-\frac{g_A^2}{64} \frac{\partial K_0}{\partial \w}(0,t) 
    (-10 M_\pi^2 t^2 +32 \w^2 M_\pi^4 -24 \w^2 M_\pi^2 t +4 \w^2 t^2 +32 M_\pi^4 t 
    \no\\&&\hspace{2.7cm}-32 M_\pi^6+t^3)\no\\&&
+\frac{1}{48} I_0(t) \Big[ 2 g_A^2 (t^2 -4 M_\pi^2 \w^2 +4 M_\pi^4 +4 \w^2 t -5 M_\pi^2 t)
    \no\\&&\hspace{1.7cm}-8 M_\pi^2 t +16 M_\pi^4 +t^2 -16 \w^2 M_\pi^2 +4 \w^2 t\Big]\no\\&&
+\frac{1}{2304 \w^3 \pi^2} \Big[g_A^4 (-48 \pi M_\pi^7-12\pi M_\pi^3 \w^2 t -24 \pi M_\pi^5 \w^2 
    +48 \pi M_\pi^5 t -12 \pi M_\pi^3 t^2 
    \no\\&&\hspace{2.7cm}+96 \w^7+4 \w^3 t^2 +40 t \w^5 +72 \pi \w^4 M_\pi^3 
    -36 \w M_\pi^4 t +6 \w M_\pi^2 t^2  
    \no\\&&\hspace{2.7cm}-12 \w^3 M_\pi^2 t +32 M_\pi^4 \w^3+48 \w M_\pi^6 -176 \w^5 M_\pi^2)
    \no\\&&\hspace{2.2cm}+g_A^2 (7 \w^3 t^2 +72 \pi M_\pi^3 \w^2 t -144\pi M_\pi^5 \w^2 
    +72 \pi \w^4 t M_\pi  
    \no\\&&\hspace{2.7cm}-456 \w^5 M_\pi^2+264 \w^3 M_\pi^4 -94 \w^3 t M_\pi^2 +192 \w^7 
    +76 \w^5 t 
    \no\\&&\hspace{2.7cm}+216 \pi M_\pi^3 \w^4+192\pi M_\pi^3 \w^2 m (c_3-c_4)(-t-2 \w^2 +2 M_\pi^2))
    \no\\&&\hspace{2.2cm}+2 \w^3 (-10 t M_\pi^2 -24 \w^2 M_\pi^2 +t^2
    +24 M_\pi^4 +4 \w^2 t ) \Big]~, \\
m F_\pi^4 h^-(\w,t) & = &  
-\frac{g_A^2}{12 \w^3} J_0(\w) (\w^2-M_\pi^2)(-2 M_\pi^2 g_A^2 +3 \w^2 +8 \w^2 m c_4)\no\\&&
+\frac{g_A^2}{12 \w^3} J_0(-\w) \Big[2 g_A^2(-3 M_\pi^4 +2 \w^4 + M_\pi^2 t + M_\pi^2 \w^2)
    \no\\&&\hspace{2.7cm}+\w^2 (\w^2 -3 M_\pi^2 +8 m c_4 (\w^2-M_\pi^2))\Big]\no\\&&
-\frac{g_A^2}{12 \w^2} \frac{\partial J_0}{\partial\w}(-\w) \Big[g_A^2 (-8 M_\pi^2 \w^2 
    -M_\pi^2 t +\w^2 t +4 \w^4 +4 M_\pi^4)
    \no\\&&\hspace{2.7cm}+2 \w^2 (\w^2-M_\pi^2) \Big]\no\\&&
-\frac{g_A^2}{8} K_0(0,t) \w (-4 M_\pi^2 +t)
+\frac{g_A^2}{32} \frac{\partial K_0}{\partial\w}(0,t) (8 M_\pi^4 -6 M_\pi^2 t +t^2)\no\\&&
-\frac{1}{24} I_0(t) \Big[2 g_A^2 (-5 M_\pi^2 +2 t)
    +(4 M_\pi^2 -t) (1 +4 m c_4) \Big]\no\\&&
-\frac{1}{1152\w^3 \pi^2} \Big[ g_A^4 (96\pi M_\pi^5 -64 \w^3 M_\pi^2 +12 \w t M_\pi^2 
    +8 \w^3 t -24 \pi M_\pi^3 t 
    \no\\&&\hspace{3.3cm}-48 \pi M_\pi^3 \w^2 +64 \w^5)
    \no\\&&\hspace{2.3cm}+g_A^2 (-18 \w^3 M_\pi^2 +48 \w^5 +36\pi \w^4 M_\pi +11 \w^3 t 
    \no\\&&\hspace{3.3cm}+16\w^3 m c_4 (15 M_\pi^2 -4 \w^2))
    \no\\&&\hspace{2.3cm}+2 \w^3 (6 M_\pi^2-t +4 m c_4 (6 M_\pi^2 - t))\Big]~.
\eea
We have used the loop functions of ref.\cite{moj}.  The $\bar{e}_i$
are scale-independent (using the same procedure to eliminate the chiral
logarithms as detailed in (I) for the $\bar{d}_i$).
It is important to stress that
we have obtained a more precise representation of the imaginary parts as compared
to~(I) since in that paper, only the leading terms were included. Here, we also have
the next--to--leading order corrections of the unitarity corrections. 
Clearly, unitarity is perturbatively fulfilled, i.e. 
${\rm Im}~T^{(4)} \sim ({\rm  Re}~T^{(2)})^2$ in a highly
symbolic notation, where $T^{(n)}$ refers to the chiral representation of the
$\pi$N amplitude to $n^{\rm th}$ order. 
It goes without saying that we also expect the
corresponding real parts to be given more accurately. 

\subsection{Counterterm combinations}
\label{sec:cts}
In the counterterm and loop contributions given above, a set
of LECs appears.  Most of these only enter in certain combinations
and some only lead to quark mass renormalizations of the 
dimension two LECs. Therefore, it is instructive to work out  how
many independent local contact terms can contribute to $\pi$N
scattering to fourth order. This can be most easily done based on a 
dispersive analysis by counting the number of possible subtractions. 
For that, it is most appropriate to describe the pertinent T--matrix
in terms of the standard invariant amplitudes $A$ and $B$,
\beq
T^\pm_{\pi N} = A^\pm + \barr{q} \, B^\pm ~,
\eeq
in a highly symbolic notation. The invariant amplitudes are functions
of two variables, which one can choose to be $\nu$ and $t$; these
count as ${\cal O}(q)$ and ${\cal O}(q^2)$, respectively. 
The most general polynom for the four amplitudes $A^\pm, B^\pm$
to fourth order commensurate with crossing and the other symmetries thus takes 
the form 
\bea
A^+_{\rm pol} &=& a_1^+ + a_2^+ t + a_3^+ \nu^2 + a_4^+ t^2 + a_5^+ t \nu^2 +
a_6^+ \nu^4~, \nonumber \\
A^-_{\rm pol} &=& \nu \, (a_1^- + a_2^- t + a_3^- \nu^2 )~, \nonumber \\
B^+_{\rm pol} &=& b_1^+ \nu~, \nonumber \\
B^-_{\rm pol} &=& b_1^- + b_2^- t + b_3^- \nu^2~.
\eea
Certain  combinations of dimension two, three and four LECs are related 
to the subtraction constants $(a_1^+, \ldots ,b_3^-)$. We refrain from 
giving the precise relationship here since it is not needed in what follows.
Therefore, in total we have 14 LECs since at third order there is one
more related to the Goldberger--Treiman discrepancy, i.e. a local
term with a LEC which allows
to express the axial--vector coupling $g_A$ in terms of the
pion--nucleon coupling $g_{\pi N}$, i.e.
\beq
g_{\pi N}= {g_A \, m\over F_\pi} \biggl( 1 - {2M_\pi^2\, \bar{d}_{18}\over g_A}
\biggr) \,\,.
\eeq
This term is important if one wants to properly account for the Born terms
expressed as a function of the pion--nucleon coupling constant.
If one then calculates to orders $q^2$, $q^3$ and
$q^4$, one has to pin down 4, 9 and 14 LECs, respectively. This
pattern is quite different from the total number of terms in the
Lagrangian allowed at
the various orders (7, 23, and 118, respectively); 
it is a general rule that simple processes
do not involve an exorbitant number of LECs. Indeed, the terms 
proportional to the dimension four LECs $\bar{e}_i$ ($i =
19,20,21,22,35,36,37,38$) only amount to quark mass renormalizations of
the dimension two LECs $c_i$ ($i=1,2,3,4$) via
\beqa\label{ce}
\tilde{c}_1 &=& c_1 -2 M^2 (\bar{e}_{22}-4 \bar{e}_{38})~,\no\\
\tilde{c}_2 &=& c_2 +8 M^2 \biggl(\bar{e}_{20}+\bar{e}_{35}
                   -\frac{g_A \bar{d}_{16}}{8 m}\biggr)~,\no\\
\tilde{c}_3 &=& c_3 +4 M^2 (2 \bar{e}_{19}- \bar{e}_{22}- \bar{e}_{36})~, \no\\
\tilde{c}_4 &=& c_4 +4 M^2 (2 \bar{e}_{21} - \bar{e}_{37})~.
\eeqa
We have also used these parameters in the one--loop graphs of order
$q^4$, although this induces some higher order
contributions. This is a very general phenomenon of CHPT calculations
at higher orders (for a discussion, see e.g. ref.\cite{BM}). 
There are different ways of determining the LECs. As in~(I), 
we use data from the physical region for doing so. Our
first strategy is to fit the renormalized $c_i$, called $\tilde{c}_i$ here,
together with the four (combinations of) dimension three LECs
$\bar{d}_1+\bar{d}_2, \bar{d}_3,\bar{d}_5, \bar{d}_{14}-\bar{d}_{15}$ 
and $\bar{d}_{18}$ and the genuine
dimension four LECs $\bar{e}_{14}, \bar{e}_{15}, \bar{e}_{16},
\bar{e}_{17}, \bar{e}_{18}$.
As enumerated before, we thus have 14 free parameters. In such
a fit, we cannot disentangle the $\tilde{c}_i$ into their quark mass 
dependent and independent pieces without further information from
other processes. This defines our best fit.
To study the convergence compared to the lower order calculations, 
we also show the best fit from~(I) and a best second order fit based
on tree diagrams with the dimension two insertions $\sim c_i$. 
One can argue that the second order contribution is given by the
amplitude up to 
second order, with the $c$'s taking on their values as given by the best
fit at that order. 
By including the amplitude at third order, the values of the $c$'s will
change,
but these changes are considered to be effects of third order. 
(The same is valid of course for the $c$'d and $d$'s, when going from
third to fourth order.)
One might also be interested in how big the contribution from genuine 
second and third order terms are (terms really proportioanl to $q^2$, 
respectively to $q^3$).
In order to address this question, we 
consider an alternative strategy, in which we fix the $c_i$ ($i=1,2,3,4$)
to their values determined from the best fit up to third order  
and use the four combinations of dimension four
LECs appearing in eq.(\ref{ce}) as fit parameters. Of course, this leads to the
same number of free parameters, but this second method allows for a clean
separation of the contributions from the various orders.
Clearly, physical observables do not depend on this
reshuffling of fit parameters (modulo higher order corrections effectively
included when using the $\tilde{c}_i$ in the loop graphs).

\section{Results}
\setcounter{equation}{0}

\subsection{The fitting procedure}

There are various possibilities to fix the LECs, a general discussion is
given in~(I). We proceed here along the same lines as in~(I), namely we fit to the
phase shifts given by three different partial wave analyses in the low
energy region. As input we use the phase shifts of the 
Karlsruhe (KA85) group~\cite{koch}, from the analysis of
Matsinos~\cite{mats} (EM98)
and the solution called SP98 from the VPI/GW
group~\cite{SAID}.~\footnote{In the meantime, novel solutions like SM99
  have appeared. Since these are not very different from  SP98 and
  we want to have a direct comparison with the results of~(I), we use
  SP98 here. We come back to this later.}
In contrast to what was done in~(I), we do not assign a common error of
3\% to the Karlsruhe and VPI phases, but rather mimic the uncertainties
of the Matsinos analysis in all cases, which is 1.5\% for $S_{31}$,
0.5\% for $S_{11}$, 1\% for $P_{33}$ and 3.5\% for the other P--waves.
This assignment gives more weight to the better determined larger
partial waves and is more natural than one common global
error.\footnote{Of course, this might induce some mismatch in the
  sense that real errors associated to the KA and VPI/GW phases are
  different from the ones of the Matsinos analysis. We believe,
  however, that this procedure is preferable to the one using common
  global errors.}  The
LEC  $\bar{d}_{18}$ is fixed by means of the Goldberger--Treiman discrepancy, i.e.
by the value for the pion--nucleon coupling constant extracted in the various
analyses. The actual values of $g_{\pi N}$ are $g_{\pi N} = 13.4\pm
0.1 \,, \,\,\, 13.18 \pm 0.12\, ,\,\,\, 13.13\pm 0.03 \, ,$
for KA85, EM98 and SP98. Throughout, we use $g_A = 1.26$, $F_\pi =
92.4\,$MeV, $m = 938.27\,$MeV and $M_\pi = 139.57\,$MeV. 
Finally, we remark that we do not use the value of the pion--nucleon 
$\sigma$--term in the fitting procedure. This has two reasons: First, as
noted before, we only want to use information from the physical
region to pin down the LECs and second, it is known that the convergence
of the chiral series for this quantity is slow~\cite{BM}. Before presenting
the results of the actual fits, we already anticipate that the EM98 data
basis will lead to the smallest $\chi^2$ for the following reasons. First,
this data base is only covering the low--energy region of pion--nucleon scattering.
Also, the representation is available on a
denser grid of points in momentum transfer. In contrast, the KA85 and SP98
analyses span a much larger range of energy and thus uncertainties also from
higher energies will play a role in the energy range considered here.
Furthermore, in~(I) we had already discussed that the extraction of the
threshold parameters from the SP98 analysis is not unproblematic. Note, however,
that the model underlying the EM98 analysis should not be used in the
unphysical region, quite in contrast to the dispersion theoretical approach
on which the KA85 phase shifts are based.

\subsection{Phase shifts and threshold parameters}

After the remarks of the preceding paragraph, we can now present results.
For the KA85 case, we have fitted the data up to 100~MeV pion  lab
momentum (i.e. 4 points per partial wave at $q_\pi = 40, 60, 79,
97$~MeV). For the analysis of Matsinos, we use 17 points for each 
partial wave in the range of $q_\pi = 41.4 - 96.3\,$~MeV. 
For the VPI SP98 solution, 
we use the 5 data points in the range between 60 and
100~MeV, which give a stable fit. Of course, we could now extend the fits to higher
energies than it was done in~(I), but for a better comparison we do not show the
results of these extended fits here. As discussed in
section~\ref{sec:cts}, we have two options for pinning down the LECs.
Using strategy one, i.e. working with the $\tilde{c}_i$,
we call the fits corresponding to the
Karlsruhe, Matsinos and VPI analysis, fit~1, 2 and 3, in order. 
The resulting LECs  are given in table~\ref{tab:LEC}. 
\renewcommand{\arraystretch}{1.2}
\begin{table}[hbt]
\begin{center}
\begin{tabular}{|c|c|c|c|}
    \hline
    LEC      &  Fit 1  &  Fit 2  & Fit 3  \\
    \hline\hline
$\tilde{c}_1$ & $-2.54\pm 0.03 $ & $-0.27\pm 0.01 $ & $-3.31\pm 0.03$ \\
$\tilde{c}_2$ & $ 0.60\pm 0.04 $ & $ 3.29\pm 0.03 $ & $ 0.13\pm 0.03$ \\
$\tilde{c}_3$ & $-8.86\pm 0.06 $ & $-1.44\pm 0.03 $ & $-10.37\pm 0.05$ \\
$\tilde{c}_4$ & $ 2.80\pm 0.13 $ & $ 3.53\pm 0.08 $ & $ 2.86\pm 0.10$ \\
\hline
$\bar{d}_1+\bar{d}_2$ 
      & $ 5.68\pm 0.09 $ & $ 4.45\pm 0.05 $ & $ 5.59\pm 0.06$ \\
$\bar{d}_3$ 
      & $-4.82\pm 0.09 $ & $-2.96\pm 0.05 $ & $-4.91\pm 0.07$ \\
$\bar{d}_5$ 
      & $-0.09\pm 0.06 $ & $-0.95\pm 0.03 $ & $-0.15\pm 0.05$ \\
$\bar{d}_{14}-\bar{d}_{15}$ 
      & $-10.49\pm 0.18 $ & $-7.02\pm 0.11 $ & $-11.14\pm 0.11$ \\
$\bar{d}_{18}$ 
      & $-1.53\pm 0.17 $ & $-0.97\pm 0.11 $ & $-0.85\pm 0.06$ \\
\hline
$\bar{e}_{14}$ & $ 6.39\pm 0.27$ & $-4.68\pm 0.14$  & $7.83 \pm 0.23$ \\
$\bar{e}_{15}$ & $ 4.65\pm 0.31$ & $-18.41\pm 0.15$ & $9.72 \pm 0.25$ \\
$\bar{e}_{16}$ & $ 7.05\pm 0.30$ & $ 7.79\pm 0.15$  & $6.42 \pm 0.25$ \\
$\bar{e}_{17}$ & $14.88\pm 0.98$ & $-17.79\pm 0.53$ & $5.47 \pm 0.64$ \\
$\bar{e}_{18}$ & $-9.15\pm 0.98$ & $ 19.66\pm 0.53$ & $-0.17\pm 0.64$ \\
\hline\hline
  \end{tabular}
  \caption{Values of the LECs in GeV$^{-1}$, GeV$^{-2}$ and
    GeV$^{-3}$ for the $\tilde{c}_i$, $\bar{d}_i$ and $\bar{e}_i$, 
    respectively, for the various fits based on the first procedure
    as described in the text.\label{tab:LEC}}
\end{center}\end{table}
\noindent
Note that the error on the LECs is purely 
the one given by the fitting routine and is certainly underestimated.
We remark that the $\tilde{c}_i$ and $\bar{d}_i$ (or combinations thereof) are
mostly of natural size, whereas some of the $\bar{e}_i$ come out fairly
large. Also, there is some sizeable variation in the actual values
of most LECs among the different fits.
The resulting S-- and P--wave phase shifts are shown in figs.~1~(fit~1),
2~(fit~2) and 3~(fit~3), in order. The corresponding $\chi^2$/dof is 0.40,
0.008 and 0.44 for fits~1,2 and 3, respectively. 
The description of the phase shifts is excellent for fit~2.
For fits~1 and 3, the $S_{31}$ and $P_{11}$ partial waves at pion
momenta above 150~MeV are somewhat off. In all cases, the description
of the $P_{33}$ partial wave is improved as compared to the third order calculation.
Since we do not fit data below $q_\pi = 40, 41, 60\,$MeV (fit~1,2,3) ,
the threshold parameters are now predictions. These are
shown for the various fits in table~\ref{tab:thr}, in comparison to the
empirical values of the various analyses. First, we observe that the
numbers resulting from the one--loop calculation are consistent
with the ``empirical'' ones. 
\begin{table}[H]
\begin{center}
\begin{tabular}{|c||c|c|c||c|c|c|}
    \hline
   Obs.    &  Fit 1  &  Fit 2  & Fit 3  & KA85 & EM98 & SP98 \\
    \hline\hline
$a^+_{0+}$  & $-0.96$ & $0.45$ & $0.27$  
      & $-0.83$ & $ 0.41\pm 0.09 $ & $ 0.0 \pm 0.1$ \\
$b^+_{0+}$  & $-5.31$ & $-4.82$ & $-7.16$  
      & $-4.40$ & $ -4.46  $ & $-4.83 \pm 0.10$ \\
$a^-_{0+}$  & $9.03$ & $7.71$ & $8.67$  
      & $ 9.17$ & $ 7.73\pm 0.06 $ & $ 8.83 \pm 0.07$ \\
$b^-_{0+}$  & $ 1.50$ & $ 1.78$ & $ 1.15$  
      & $ 0.77$ & $ 1.56 $ & $ 0.07 \pm 0.07$ \\
$a^+_{1-}$  & $-5.66$ & $-5.87$ & $-4.81$  
      & $-5.53$ & $-5.46\pm 0.10 $ & $-5.33 \pm 0.17$ \\ 
$a^+_{1+}$  & $ 13.15$ & $13.04$ & $ 13.18$  
      & $ 13.27$ & $ 13.13\pm 0.13 $ & $ 13.6 \pm 0.1$ \\
$a^-_{1-}$  & $-1.25$ & $-1.17$ & $-0.79$  
      & $-1.13$ & $-1.19\pm 0.08 $ & $-1.00 \pm 0.10$ \\
$a^-_{1+}$  & $-7.99$ & $-8.21$ & $-7.54$  
      & $-8.13$ & $-8.22\pm 0.07 $ & $-7.47 \pm 0.13$ \\
  \hline\hline
  \end{tabular}
  \caption{Values of the S-- and P--wave threshold parameters for the various fits
           as described in the text in comparison to the respective
           data. Note that we have extracted $b_{0+}^\pm$ from the
           Matsinos phase shifts and thus no uncertainty is given. 
           Units are appropriate inverse powers of the pion mass times 10$^{-2}$.
           \label{tab:thr}}
\end{center}
\end{table}
\noindent
The agreement with the threshold parameters based on the chiral amplitude
with the ones based on the approaches underlying the various partial wave
analyses is best for fit~2, slightly worse for fit~1 and clearly problematic
for some of the parameters of fit~3. The reason for this was already spelled
out in~(I). The bands for the S--wave scattering lengths  $a^+_{0+}$ and 
$a^-_{0+}$ are as in~(I) since the uncertainties extracted there are
mostly due to the input and not to the  theory. They agree with the
recent determinations from the shift and width of pionic hydrogen and
deuterium, cf. Fig.2 in ref.\cite{PSI}. For comparison, we translate
our bands on the isoscalar and isovector scattering lengths into
the physical ones,
\beqa
a_{\pi^- p \to \pi^0 n} &=& -0.131 \ldots -0.117 \, M_{\pi}^{-1} \quad 
                            [(-0.128 \pm 0.006) \, M_{\pi}^{-1}]~,
                            \nonumber\\
a_{\pi^- p \to \pi^- p} &=& 0.073 \ldots 0.093 \, M_{\pi}^{-1} \quad 
                            [(0.0883 \pm 0.0008) \, M_{\pi}^{-1}]~,
\eeqa
where the experimental numbers (in the square brackets) are taken from ref.\cite{PSI}. 
 Note, however, that recent
progress in calculating $\pi^- p$ atoms in effective field theory lets
one expect that the uncertainty due to electromagnetic corrections for
the band derived from the hydrogen shift has been underestimated, see
e.g. ref.\cite{bern}. Furthermore, only recently deuteron
wave functions have been obtained precisely enough in an EFT approach to
readdress the question of the deuterium shift
constraining the elementary $\pi N$ amplitudes. 
It would also be worthwhile to
repeat the EFT calculation of pion--deuteron scattering~\cite{silas}
using our fourth order $\pi N$ amplitudes as input.
As argued before, we can study the convergence of the chiral expansion.
In figs.1--3, the  dot--dashed, dotted, dashed  and solid lines refer
to the best fits up to first, second, third and fourth order, respectively.
Since we used the errors of the Matsinos analysis, it is best to consider fit~2
shown in fig.~2. In most cases, the fourth order corrections are 
smaller than the third
order ones, indicating convergence. This could not be concluded from the
third order calculation, compare the discussion in~(I) and
ref.\cite{moj}. Note also that in some partial waves the second
order result is close to the data. The resulting values of the
$c_i$ are very different from the ones given in 
table~\ref{tab:LEC2}. The second order best fit based on the (KA85,
EM98, SP98) analysis leads to $c_1 = (-0.81, -0.77,-1.06)$, 
$c_2 = (2.47,2.69,2.36)$, $c_3 = (-3.78,-3.96,-4.04)$ and
$c_4 = (2.49,2.64,2.35)$ (all in GeV$^{-1}$).\footnote{Note that for 
such a second order fit the differentiation between the $c_i$ and 
$\tilde{c}_i$ becomes meaningless.} That these values are
very different from the ones based on a one--loop third order
amplitude fit was already pointed out in ref.\cite{bkmppn}. 
It is of particular interest to study the convergence of the S--wave
scattering lengths, which has been already discussed in
ref.\cite{bkmpin1,bkmpin2} estimating LECs from resonance saturation. Our
results are summarized in table~\ref{tab:aconv1}.  Although it was
already shown in ref.\cite{bkmpin2} that there are no fourth order
corrections to $a_{0+}^-$, the readjustment of the LECs when going
from third to fourth order leads to a small difference. That this
difference is so small is also in agreement with ref.\cite{bkmpin2},
where it was argued that the dominant correction to the Weinberg--Tomozawa
low--energy theorem is a pion loop effect. For fits 1 and 2, the
fourth order correction to the isoscalar S--wave scattering length is
fairly small, even for fit~3 the dominant correction is the one from
second to third order. 
\begin{table}[htb]
\begin{center}
\begin{tabular}{|r|r|r|r|r|}
    \hline
          & ${\cal O}(q)$  &   ${\cal O}(q^2)$  &  ${\cal O}(q^3)$ 
          & ${\cal O}(q^4)$  \\
    \hline\hline
           fit~1      & 0.0  & 0.46  & $-$1.00  & $-$0.96 \\
$a_{0+}^+ \quad$fit~2 & 0.0  & 0.24  &    0.49  &    0.45 \\
           fit~3      & 0.0  & 1.01  &    0.14  &    0.27 \\ 
\hline
           fit~1      & 7.90 & 7.90  &    9.05  &    9.03 \\
$a_{0+}^- \quad$fit~2 & 7.90 & 7.90  &    7.72  &    7.71 \\
           fit~3      & 7.90 & 7.90  &    8.70  &    8.67 \\
\hline\hline
  \end{tabular}
  \caption{Convergence of the S--wave scattering lengths for the
   three fits. ${\cal O}(q^n)$ means that all terms up-to-and-including
   order $n$ are given. Units are 10$^{-2}/M_\pi$.
    \label{tab:aconv1}}
\end{center}\end{table}

The second option is to keep the dimension two LECs fixed to their value 
determined from the third order fits and fit the additional four dimension four
combinations. This allows for a clean discussion of the various contributions
to the chiral expansion. The results for the LECs are shown in 
table~\ref{tab:LEC2}.\footnote{The slight differences for the values of the $c_i$
  as compared to the ones given in~(I) stem from the fact that we use
  different error bars for the KA85 and SP98 partial waves as explained before.}
\begin{table}[h]
\begin{center}
\begin{tabular}{|c|c|c|c|}
    \hline
    LEC      &  Fit 1*  &  Fit 2*  & Fit 3*  \\
    \hline\hline
${c}_1$ & $-1.21*$ & $-1.42*$ & $-1.47*$ \\
${c}_2$ & $ 3.29*$ & $ 3.13*$ & $ 3.26*$ \\
${c}_3$ & $-5.91*$ & $-5.85*$ & $-6.14*$ \\
${c}_4$ & $ 3.47*$ & $ 3.50*$ & $ 3.50*$ \\
\hline
$\bar{d}_1+\bar{d}_2$ 
      & $ 5.32\pm 0.08 $ & $ 5.26\pm 0.05 $ & $ 4.90\pm 0.05$ \\
$\bar{d}_3$ 
      & $-4.37\pm 0.09 $ & $-3.61\pm 0.05 $ & $-4.19\pm 0.07$ \\
$\bar{d}_5$ 
      & $-0.13\pm 0.06 $ & $-1.03\pm 0.03 $ & $-0.16\pm 0.05$ \\
$\bar{d}_{14}-\bar{d}_{15}$ 
      & $-9.31\pm 0.18 $ & $-8.70\pm 0.11 $ & $-9.31\pm 0.10$ \\
$\bar{d}_{18}$ 
      & $-1.47\pm 0.16 $ & $-1.49\pm 0.10 $ & $-0.84\pm 0.06$ \\
\hline
$\bar{e}_{14}$ & $ 2.33\pm 0.27$ & $ 2.64\pm 0.14$  & $4.19 \pm 0.23$ \\
$\bar{e}_{15}$ & $-2.21\pm 0.30$ & $-3.33\pm 0.15$  & $4.54 \pm 0.25$ \\
$\bar{e}_{16}$ & $ 5.69\pm 0.28$ & $ 4.02\pm 0.14$  & $2.74 \pm 0.24$ \\
$\bar{e}_{17}$ & $ 6.18\pm 0.98$ & $ 5.14\pm 0.53$  & $7.20 \pm 0.64$ \\
$\bar{e}_{18}$ & $-1.27\pm 0.98$ & $-2.56\pm 0.53$  & $-3.36\pm 0.64$ \\
\hline
$\bar{e}_{22}-4 \bar{e}_{38}$ 
               & $15.06\pm 0.79$ & $ 7.38\pm 0.40$  & $27.72 \pm 0.74$ \\
$\bar{e}_{20}+\bar{e}_{35} - g_A \bar{d}_{16}/ (8 m) $ 
               & $-15.28\pm 0.43$ & $-10.49\pm 0.21$ & $-17.35 \pm 0.36$ \\
$2 \bar{e}_{19}- \bar{e}_{22}- \bar{e}_{36} $  
               & $-3.58\pm 0.83$ & $-1.49\pm 0.40$  & $-25.12 \pm 0.69$ \\
$ 2 \bar{e}_{21} - \bar{e}_{37} $ 
               & $-7.12\pm 1.96$ & $-1.66\pm 1.10$ & $-5.00 \pm 1.43$ \\
\hline\hline
  \end{tabular}
  \caption{Values of the LECs in GeV$^{-1}$, GeV$^{-2}$ and
    GeV$^{-3}$ for the $c_i$, $\bar{d}_i$ and $\bar{e}_i$, 
    respectively, for the various fits based on the second procedure
    as described in the text. The $*$ denotes an input quantity.
    \label{tab:LEC2}}
\end{center}\end{table}
\noindent
The stability of the values for most of the LECs is better than
in the previous case and the fourth order LECs $\bar{e}_i$,
$i=14,\ldots,18$ smaller (more natural). Some of the additional
combinations of dimension four LECs are fairly large and vary
considerably for the various fits. 
The resulting S-- and P--wave phase shifts are shown in figs.~4~(fit~1*),
5~(fit~2*) and 6~(fit~3*), in order. The corresponding $\chi^2$/dof is 0.50,
0.14 and 0.58 for fits~1*,2* and 3*, respectively. In these plots a different 
way of looking at the convergence properties of the amplitude is adopted:
all four curves are based on the same fit and are thus obtained with the same set of LECs.
The  dot--dashed, dotted, dashed  and solid lines show
the contributions from the amplitude up to first, second, 
third and fourth order, respectively.
We remark that the
fourth order contributions are mostly small, with the exception of the
$P_{11}$ and $P_{13}$ partial waves. 
The threshold parameters determined from these fits come out very close
to the values given before and we thus refrain from adding another table
here. Similar remarks hold for the convergence of the S--wave
scattering lengths, only that in this way of fixing the LECs there is
indeed no contribution to the isovector one from fourth order.

\subsection{Sigma term, subthreshold parameters and further comments}

In this section, we briefly discuss the so--called subthreshold
parameters and the sigma term. Already in~(I) we noted that the
representation of the chiral amplitude, when pinned down by scattering 
data, is not very precise in the unphysical region. In particular,
the small isoscalar amplitudes are obtained from various contributions,
which are individually much larger than their sum. Consequently, this
fine balance which is enforced through the fit in the physical region
down to the scattering lengths is disturbed because the strict $1/m$
expansion  performed here does not properly account for all cuts
appearing in the $\pi$N amplitude. In fact, using our fourth
order representation, we do not find an 
improvement of the subthreshold parameters as given in~(I), in some
cases even a clear disimprovement. This problem could e.g. be
circumvented in the formulation of ref.\cite{BL}. Another
option is to pin down the LECs inside the Mandelstam triangle~\cite{paul},
which will lead to an improved representation in the unphysical
region. This is also reflected in the prediction for the sigma term,
which came out rather large in the fits shown in~(I), but was
considerably different (and consistent with the result from
dispersion theory) based on the method used in~\cite{paul}.

\medskip\noindent
We now consider the sigma term, which is the matrix element
of the explicit chiral symmetry breaking part of the QCD Lagrangian sandwiched
between proton states at zero momentum transfer. While at third order 
we can directly give the sigma term, $\sigma(0)$, in terms of the LEC $c_1$,
this can no longer be done at fourth order due to the appearance of the LEC
combination $2e_{115}+ e_{116}/2$. These operators contribute to the
nucleon mass shift and the sigma term (scalar form factor) as noted before. These
contact interactions have no pion matrix--elements and therefore can
not appear in the scattering amplitude, even not in higher order loop
graphs. We therefore use a more indirect method to determine the
sigma term. For that, we consider $\Sigma = F_\pi^2 \bar{D}^+ (\nu =0,
t = 2M_\pi^2)$ which can be related to $\sigma (0)$ by the venerable low--energy
theorem of ref.\cite{BPP}. There exists a whole family of
relations between $\Sigma$ and certain combinations of threshold
parameters, as detailed in ref.\cite{juerg}. These relations
have been worked out to third order and should be generalized to
fourth order. We will use here the version given in ref.\cite{glls},
\beq\label{sumG}
\Sigma = \pi F_\pi^2 [(4+2\mu+\mu^2)a_{0+}^+ -4M_\pi^2 b_{0+}^+
+ 12\mu M_\pi^2 a_{1+}^+ ] + \Sigma_0~,
\eeq
with $\Sigma_0 =-12.6\,$MeV and $\mu = M_\pi/m \simeq 1/7$. 
Using the pertinent threshold parameters from the fourth (third)
order representation, we find $\Sigma = 65 \,(62)\,$MeV, $73\,
(79)\,$MeV and $90\, (82)\,$MeV for fits 1, 2 and 3, respectively.
A special variant, which also contains some fourth order
pieces, has recently been given by Olsson~\cite{MO},
\beqa\label{sumO}
{\Sigma} &=& [{F_\pi^2}\, F(2M_\pi^2)]~, \\
 F(2M_\pi^2) &=& 14.5 \, a_{0+}^{+} - 5.06\, 
(a_{0+}^{1/2})^2 - 10.13\, (a_{0+}^{3/2})^2 - 16.65\, b_{0+}^{+}
- 0.06 \, a_{1-}^{+} + 5.70 \, a_{1+}^{+} - 0.05~,\nonumber
\eeqa
with the quantities on the right--hand--side being given in units 
of the pion mass. This leads to $\Sigma = 73 \,(69)\,$MeV, $85\,
(91)\,$MeV and $104\, (93)\,$MeV for fits 1, 2 and 3, respectively.
We consider the differences between the results based on eqs.(\ref{sumG})
and (\ref{sumO}) (and also using the fourth order results for the threshold
parameters in the third order representation, eq.(\ref{sumG}))
as an indication of the size of the fourth order
terms. We note that the values we find for the Karlsruhe analysis are
consistent with the direct determination based on hyperbolic
dispersion relations~\cite{koch2}, whereas the results based on the SP98
partial waves lead to a sizeably larger value than advocated by the
VPI/GW group~\cite{pavan}.

\bigskip\bigskip

\section*{Acknowledgements}

We are grateful to Thomas Becher and Bastian Kubis for some clarifying discussions.
One of us (N.F.) would like to thank all members of the Kellogg Radiation
Lab for the hospitality extended to her during a stay while this work
was completed.

\bigskip\bigskip\bigskip\bigskip
\appendix
\section{Threshold parameters}
\def\theequation{\Alph{section}.\arabic{equation}}
\setcounter{equation}{0}
\label{app:sub}

In this appendix, we give the analytical expressions for the threshold
parameters up to fourth order. These read:

\bea
a_{0+}^+ & = & 
\frac{M_\pi^2 [-g_A^2 + 8 m (-2 c_1 +c_2+c_3)]}{16 \pi (m+M_\pi) F_\pi^2}\no\\&&
+\frac{3 g_A^2 m M_\pi^3}{256 \pi^2 (m+M_\pi) F_\pi^4}\no\\&&
-\frac{g_A^2 M_\pi^4}{64 \pi (m+M_\pi) m^2 F_\pi^2}
-\frac{4 M_\pi^4 c_1 c_2}{\pi (m+M_\pi) F_\pi^2} +\frac{2 m M_\pi^4 c_1 \ell_3}{\pi (m+M_\pi) F_\pi^4}
-\frac{g_A M_\pi^4 (2 \bar{d}_{16}-\bar{d}_{18})}{4 \pi (m+M_\pi) F_\pi^2}
\no\\&&
+\frac{2 M_\pi^4 m (2 \bar{e}_{14}+2 \bar{e}_{15} +2 \bar{e}_{16} +2 \bar{e}_{19} +2 \bar{e}_{20} +2 \bar{e}_{35} -\bar{e}_{36} -4 \bar{e}_{38} )}{\pi (m+M_\pi) F_\pi^2}
\no\\&&
-\frac{M_\pi^4[8-3 g_A^2+2 g_A^4 +4 m (2 c_1-c_3)]}{256 \pi^3 (m+M_\pi) F_\pi^4}~,
\\
a_{0+}^- & = & \frac{m M_\pi}{8 \pi (m+M_\pi) F_\pi^2} \no\\&&
+\frac{M_\pi^3 (g_A^2+32 m^2 (\bar{d}_1+\bar{d}_2+\bar{d}_3+2 \bar{d}_5))}{32 \pi m (m+M_\pi) F_\pi^2}+\frac{M_\pi^3 m}{64 \pi^3  (m+M_\pi) F_\pi^4}~,
\\
b_{0+}^+ & = & 
\frac{g_A^2 (4 m^2 +2 m M_\pi - M_\pi^2)}{64 \pi m^2 (m+M_\pi) F_\pi^2} 
+ \frac{2  c_1 (2 m M_\pi - M_\pi^2) + (c_2+c_3) (4 m^2 -2 m M_\pi + M_\pi^2)}{8 \pi m (m+M_\pi) F_\pi^2}\no\\&&
+\frac{M_\pi(g_A^2 +8 m c_2)}{16 \pi m  (m+M_\pi) F_\pi^2}
+\frac{g_A^2 M_\pi (154 m^2 -18 m M_\pi +9 M_\pi^2)}{3072 \pi^2 m (m+M_\pi) F_\pi^4}\no\\&&
-\frac{g_A^2 M_\pi^2 (-16 m^2 -2 m M_\pi + M_\pi^2)}{256 \pi (m+M_\pi) m^4 F_\pi^2}
+\frac{M_\pi^2 c_2}{2\pi (m+M_\pi) m F_\pi^2}
\no\\&&
-\frac{M_\pi^2 c_1 c_2 (4 m^2 -2 m M_\pi + M_\pi^2)}{\pi (m+M_\pi) m^2 F_\pi^2}
-\frac{M_\pi^3 c_1 \ell_3 (2m-M_\pi)}{2\pi(m+M_\pi) m F_\pi^4}
\no\\&&
-\frac{M_\pi^2 [8 m^2 (\bar{d}_{14}-\bar{d}_{15}) -g_A \bar{d}_{18} (M_\pi^2 -2 m M_\pi +4 m^2)]}{16\pi (m+M_\pi) m^2 F_\pi^2}
\no\\&&
-\frac{M_\pi^2 (\bar{e}_{14}+\bar{e}_{15}+\bar{e}_{16})(-8 m^2 +2 m M_\pi - M_\pi^2)}{\pi (m+M_\pi) m F_\pi^2}
-\frac{M_\pi^2 (2 m M_\pi - M_\pi^2) (\bar{e}_{22}-4 \bar{e}_{38})}{2 \pi (m+M_\pi) m F_\pi^2}
\no\\&&
-\frac{M_\pi^2 [2(\bar{e}_{20}+\bar{e}_{35}-\frac{g_A \bar{d}_{16}}{8 m})+(2 \bar{e}_{19}-\bar{e}_{22}-\bar{e}_{36})]
(-4 m^2+2 m M_\pi - M_\pi^2)}{2 \pi (m+M_\pi) m F_\pi^2}
\no\\&&
-\frac{M_\pi^2}{9216 \pi^3 (m+M_\pi) m^2 F_\pi^4} [-432m^2-144m M_\pi+72 M_\pi^2 +g_A^2(44m^2+54 m M_\pi-27 M_\pi^2)
\no\\&&
   +g_A^4(m^2(-88+192 \pi)-36 m M_\pi+18 M_\pi^2) +m c_1 (1248 m^2 -144 m M_\pi +72 M_\pi^2)
\no\\&&
   -24 m^3 c_2 +m c_3(-768 m^2+72 m M_\pi-36 M_\pi^2) ]~,
\\
b_{0+}^- & = & 
\frac{2 m^2 -2 m M_\pi +M_\pi^2}{32\pi m M_\pi (m+M_\pi) F_\pi^2}\no\\&&
+\frac{1-2 g_A^2}{16 \pi (m+M_\pi) F_\pi^2}\no\\&&
+\frac{M_\pi g_A^2 (-10 m^2 -2 m M_\pi +M_\pi^2)}{128 \pi m^3 (m+M_\pi) F_\pi^2}
-\frac{M_\pi c_4}{4 \pi (m+M_\pi) F_\pi^2}\no\\&&
+\frac{M_\pi  [(\bar{d}_1+\bar{d}_2+\bar{d}_3)(6 M_\pi^2 -2 m M_\pi +M_\pi^2)+\bar{d}_5(4 m^2 -4 m M_\pi +2 M_\pi^2)]}
       {4 \pi m (m+M_\pi) F_\pi^2}\no\\&&
-\frac{M_\pi (4 m^2 +6 m M_\pi -3 M_\pi^2-14 g_A^2 m^2)}{768 \pi^3 m (m+M_\pi) F_\pi^4}\no\\&&
-\frac{3 g_A^2 M_\pi^2}{64\pi (m+M_\pi) m^2 F_\pi^2}
+\frac{M_\pi^2 (\bar{d}_1+\bar{d}_2+3 \bar{d}_3 +2 \bar{d}_5 +g_A \bar{d}_{18})}{2 \pi (m+M_\pi) F_\pi^2}
\no\\&&
-\frac{M_\pi^2(18+ 69\pi g_A^2 +(4-12\pi)g_A^4)}{2304 \pi^3 (m+M_\pi)  F_\pi^4}~,
\\
a_{1-}^- & = & 
-\frac{g_A^2 m }{24\pi M_\pi (m+M_\pi) F_\pi^2}\no\\&&
+\frac{3-2 g_A^2 +8 m c_4}{48\pi (m+M_\pi) F_\pi^2}\no\\&&
+\frac{M_\pi[3 -3 g_A^2+24 m c_4 -32 m^2 (\bar{d}_1+\bar{d}_2) +16 m^2 g_A \bar{d}_{18}]}{96 \pi (m+M_\pi) F_\pi^2}\no\\&&
-\frac{M_\pi m^2[3 +g_A^2(21+36\pi)+g_A^4 (2+24 \pi)]}{3456 \pi^3 (m+M_\pi) F_\pi^4}\no\\&&
-\frac{g_A^2 M_\pi^2}{192 \pi (m+M_\pi) m^2 F_\pi^2}
-\frac{M_\pi^2 c_1}{6\pi (m+M_\pi) m F_\pi^2}
+\frac{M_\pi^2(\bar{d}_1+\bar{d}_2+3 \bar{d}_3+2 \bar{d}_5 +g_A \bar{d}_{18})}{6\pi (m+M_\pi) F_\pi^2}
\no\\&&
+\frac{2 m M_\pi^2 (2 \bar{e}_{17} +2 \bar{e}_{18}+2 \bar{e}_{21} - \bar{e}_{37})}{3\pi (m+M_\pi) F_\pi^2}
\no\\&&
-\frac{M_\pi^2}{6912\pi^3(m+M_\pi) F_\pi^4} [18+g_A^2(72-33\pi)+g_A^4(4+30\pi)
+96 \pi g_A^2 m c_3 
\no\\&&
+(176-96\pi) g_A^2 m c_4 ]~,
\\
a_{1-}^+ & = & 
-\frac{g_A^2 m}{12 \pi M_\pi (m+M_\pi) F_\pi^2}\no\\&&
-\frac{g_A^2 +2 m c_3}{12 \pi (m+M_\pi) F_\pi^2}\no\\&&
+\frac{M_\pi [c_2 +2 m (\bar{d}_{14}-\bar{d}_{15}+g_A \bar{d}_{18})]}{6 \pi (m+M_\pi) F_\pi^2}
-\frac{m M_\pi g_A^2 (231 \pi+g_A^2 (112+96 \pi))}{13824 \pi^3 (m+M_\pi) F_\pi^4}\no\\&&
+\frac{g_A^2 M_\pi^2}{192\pi(m+M_\pi) m^2 F_\pi^2}
+\frac{M_\pi^2(2 c_1-c_2-c_3)}{8\pi(m+M_\pi) m F_\pi^2}
+\frac{M_\pi^2(3( \bar{d}_{14}-\bar{d}_{15}) +2 g_A \bar{d}_{18})}{6\pi (m+M_\pi) F_\pi^2}
\no\\&&
+\frac{2 m M_\pi^2(-4 \bar{e}_{14}-2 \bar{e}_{15}-2 \bar{e}_{19} +\bar{e}_{22}+\bar{e}_{36})}{3\pi(m+M_\pi) F_\pi^2}
\no\\&&
-\frac{M_\pi^2}{6912\pi^3(m+M_\pi) F_\pi^4} [-36 +g_A^2(133-48\pi)+g_A^4(74+12\pi)
\no\\&&
-6 m (52c_1-c_2-32 c_3(1+g_A^2 \pi) +32 \pi c_4) ]~,
\\
a_{1+}^- & = & 
-\frac{m g_A^2}{24 \pi M_\pi (m+M_\pi) F_\pi^2}\no\\&&
-\frac{g_A^2 +2 m c_4}{24 \pi (m+M_\pi) F_\pi^2}\no\\&&
-\frac{M_\pi[3 g_A^2 +32 m^2 (\bar{d}_1+\bar{d}_2)-16 m^2 g_A \bar{d}_{18}]}{96 \pi m (m+M_\pi) F_\pi^2}
-\frac{M_\pi m [3 +g_A^2 (21-18 \pi) +g_A^4 (2-12 \pi)]}{3456 \pi^3 (m+M_\pi) F_\pi^4}\no\\&&
-\frac{g_A^2 M_\pi^2}{48\pi(m+M_\pi) m^2 F_\pi^2}
+\frac{M_\pi^2 c_1}{12\pi(m+M_\pi) m F_\pi^2}
+\frac{M_\pi^2 (\bar{d}_1+\bar{d}_2+3 \bar{d}_3+2 \bar{d}_5+g_A \bar{d}_{18})}{6\pi(m+M_\pi) F_\pi^2}
\no\\&&
+\frac{m M_\pi^2(-2\bar{e}_{17} -2 \bar{e}_{18}-2 \bar{e}_{21} +\bar{e}_{37})}{3\pi(m+M_\pi) F_\pi^2}
\no\\&&
+\frac{M_\pi^2}{6912 \pi^3 (m+M_\pi) F_\pi^4} [-18 +g_A^2(36+141\pi)+g_A^4(-4+42\pi)
\no\\&&
+8 m g_A^2(-12\pi c_3+(12\pi+11)c_4) ]~,
\\
a_{1+}^+ & = & 
\frac{g_A^2 m}{24 \pi M_\pi (m+M_\pi) F_\pi^2}\no\\&&
+\frac{g_A^2 -4 m c_3}{24 \pi (m+M_\pi) F_\pi^2}\no\\&&
+\frac{M_\pi[3 g_A^2 +16 m c_2 -16 m^2 (\bar{d}_{14}-\bar{d}_{15}+g_A \bar{d}_{18})]}{96 \pi m (m+M_\pi) F_\pi^2}
-\frac{M_\pi m g_A^2 (231 \pi +g_A^2 (-56+96 \pi))}{13824 \pi^2  (m+M_\pi) F_\pi^4}\no\\&&
+\frac{g_A^2 M_\pi^2}{48\pi(m+M_\pi) m^2 F_\pi^2}
-\frac{ g_A M_\pi^2 \bar{d}_{18}}{6\pi(m+M_\pi) F_\pi^2}
+\frac{2 m M_\pi^2 (-4 \bar{e}_{14}-2 \bar{e}_{15}-2 \bar{e}_{19} +\bar{e}_{22}+\bar{e}_{36})}{3\pi(m+M_\pi)F_\pi^2}
\no\\&&
-\frac{M_\pi^2}{6912 \pi^3(m+M_\pi) F_\pi^4} [-36+g_A^2(133+24\pi)+g_A^4(-10+66\pi)
\no\\&&
+6 m(-52c_1+c_2+(32-16\pi g_A^2)c_3+16\pi g_A^2 c_4)]~.
\eea

\bigskip\bigskip

\newpage

\noindent {\Large {\bf Figures}}

$\,$

\begin{figure}[H]
\centerline{
\epsfysize=7in
\epsffile{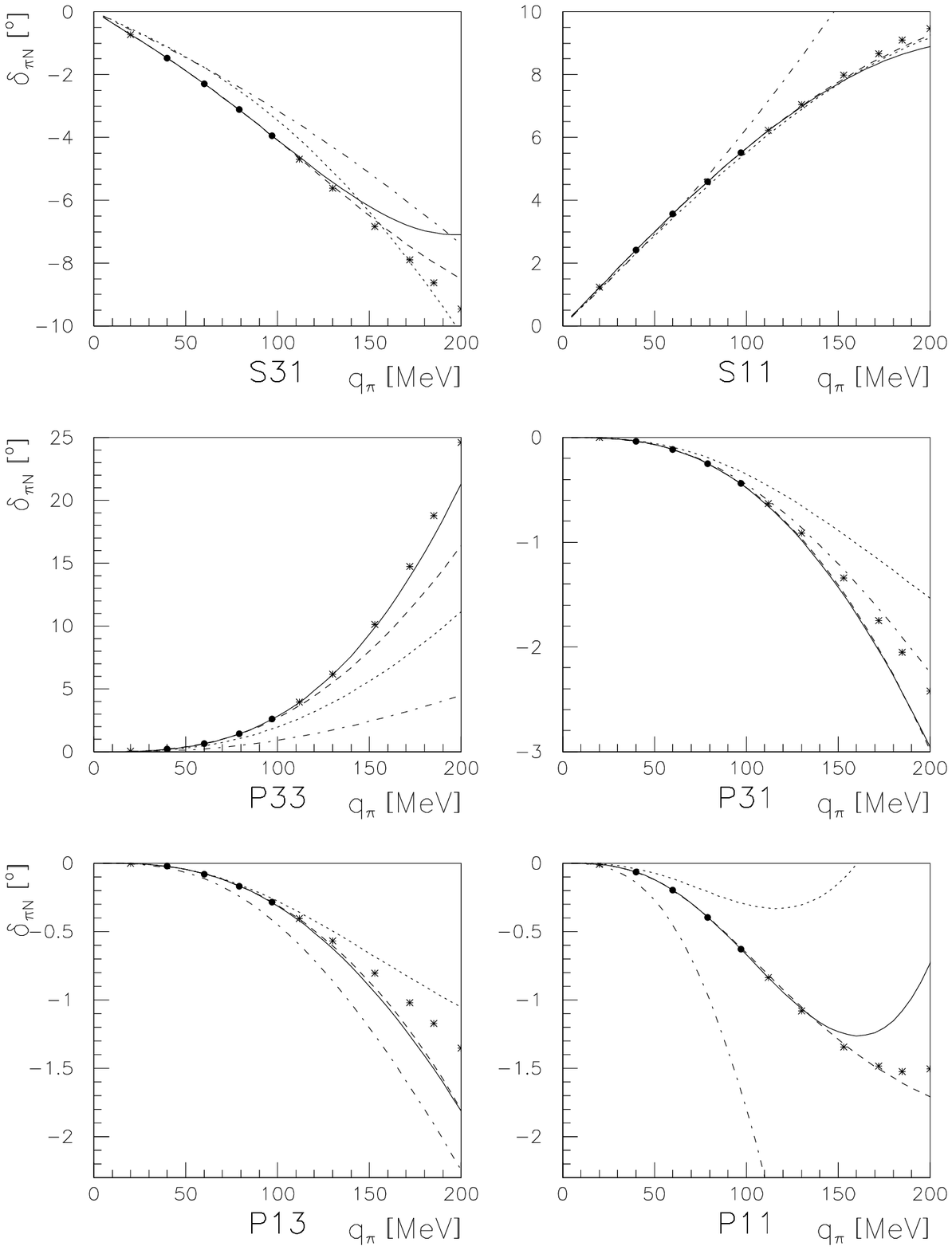}
}
\vskip 0.5cm
\caption{
Fits and predictions for the KA85 phase shifts as a
function of the pion laboratory  momentum $q_\pi$ 
using strategy one as explained in the text. 
Fitted in each partial wave are the data between 40 and 97~MeV (filled
circles). For higher and lower energies, the phases are predicted as
shown by the solid lines. The other lines refer to the best fits at the
various orders as explained in the text.}

\end{figure}

\newpage

\vskip 1cm

\begin{figure}[bht]
\centerline{
\epsfysize=7in
\epsffile{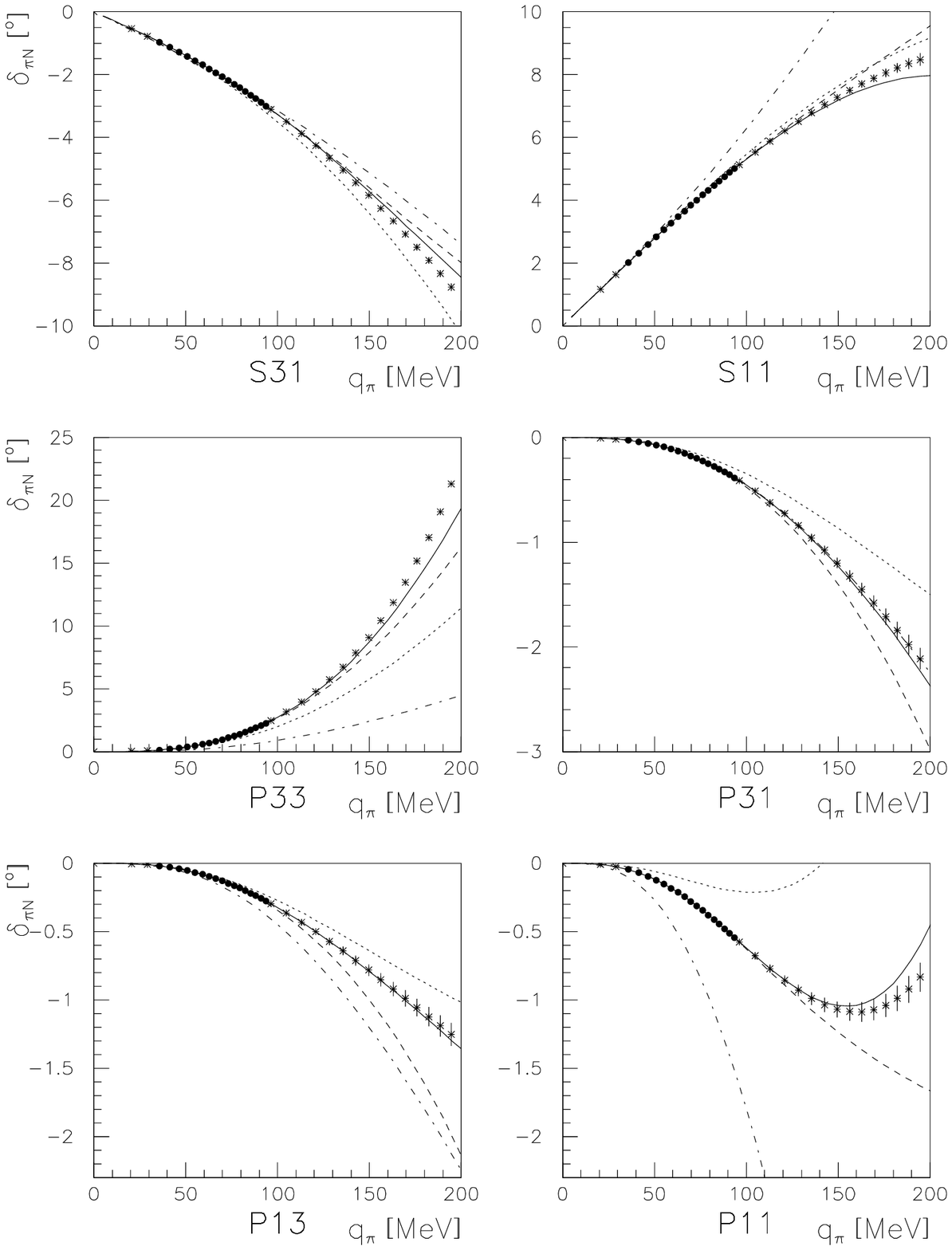}
}
\vskip 0.5cm
\caption{
Fits and predictions for the EM98 phase shifts  as a
function of $q_\pi$ using strategy one as explained in the text. 
Fitted in
each partial wave are the data between 41 and 97~MeV (filled circles). For higher
and lower energies, the phases are predicted as shown by the solid lines.
The other lines refer to the best fits at the various orders 
as explained in the text.
}

\end{figure}

\newpage

\vskip 1cm

\begin{figure}[bht]
\centerline{
\epsfysize=7in
\epsffile{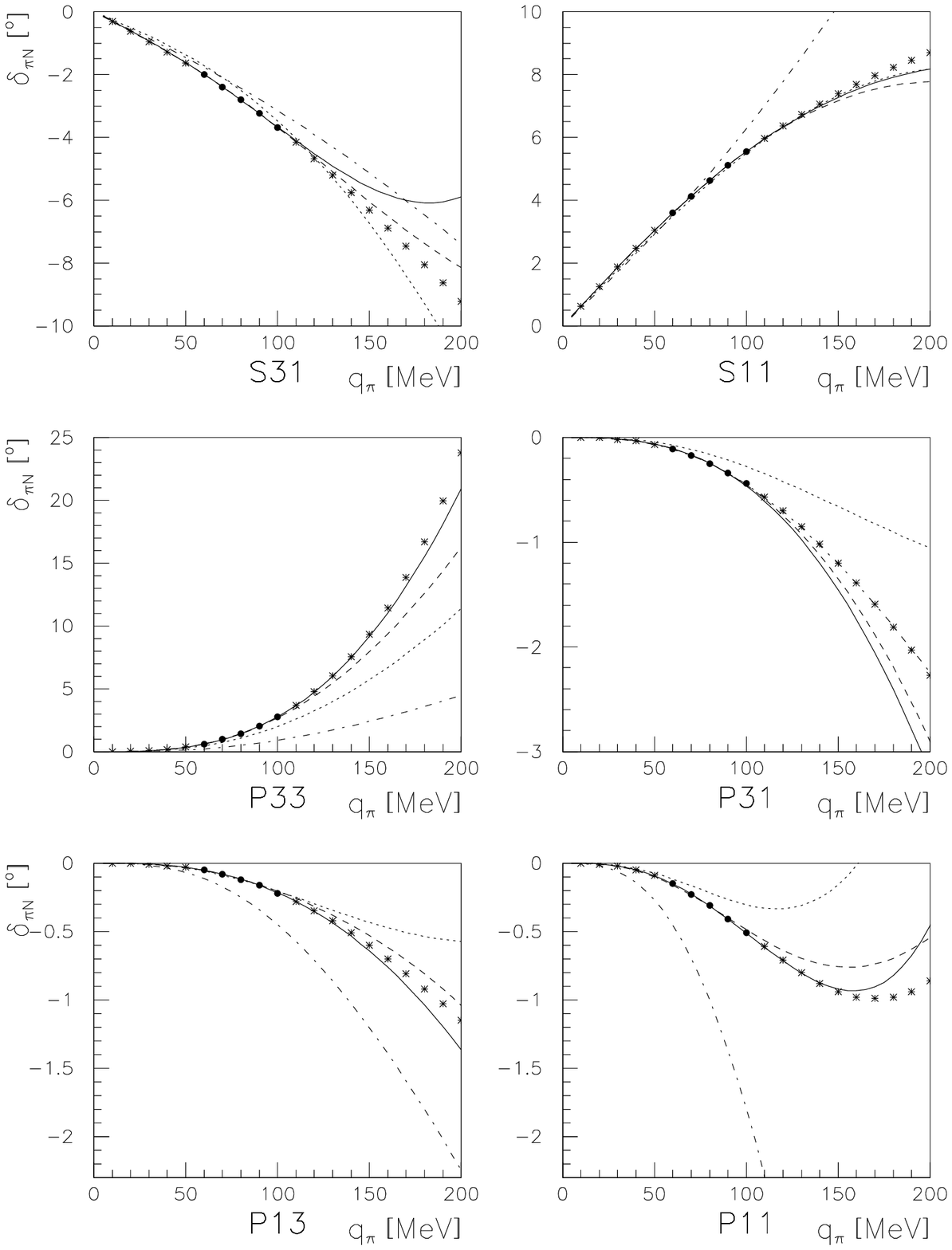}
}
\vskip 0.5cm
\caption{
Fits and predictions for the SP98 phase shifts as a
function of the pion laboratory momentum $q_\pi$
using strategy one as explained in the text. Fitted in
each partial wave are the data between 60 and 100~MeV (filled circles). For higher
and lower energies, the phases are predicted as shown by the solid lines.
The other lines refer to the best fits at the various orders as 
explained in the text.
}

\end{figure}

\vskip 1cm

\begin{figure}[bht]
\centerline{
\epsfysize=7in
\epsffile{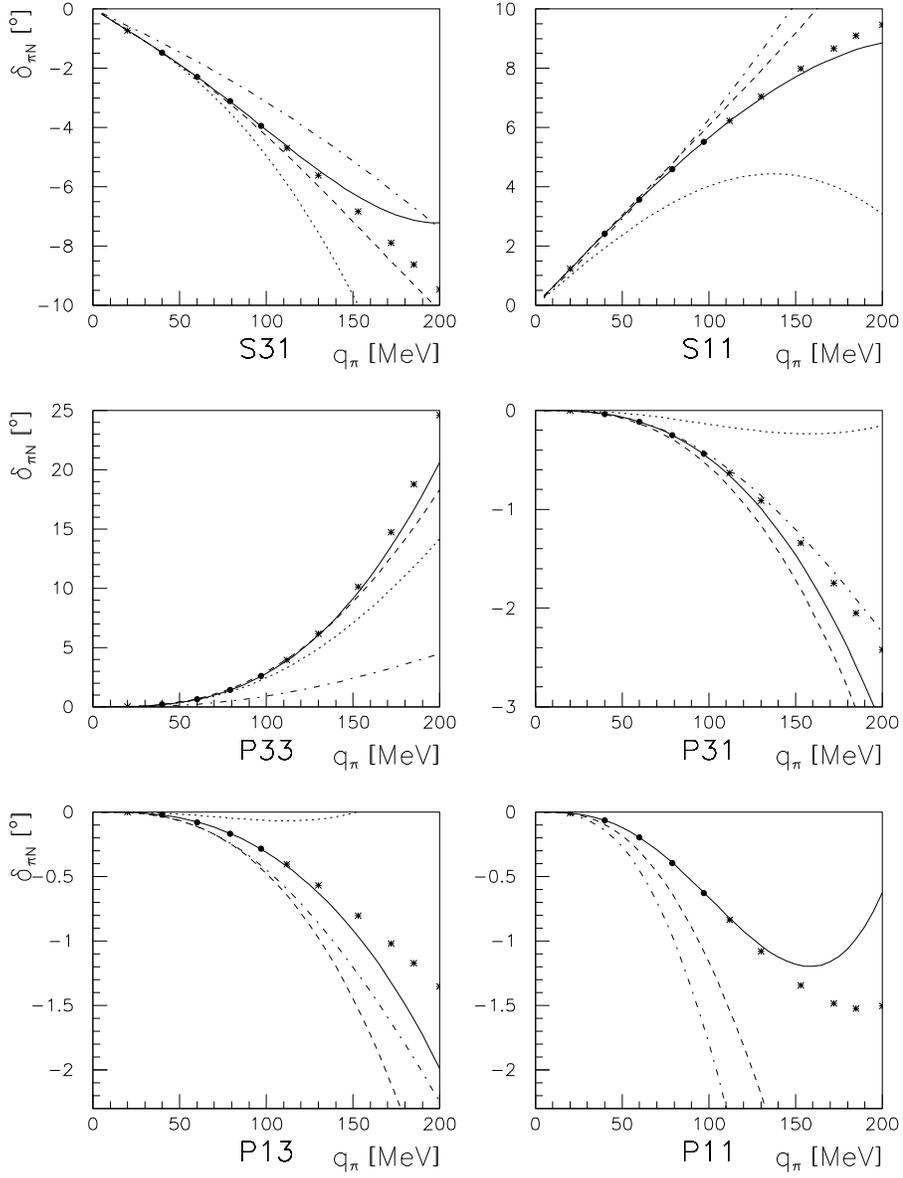}
}
\vskip 0.5cm
\caption{
Fits and predictions for the KA85 phase shifts as a
function of the pion laboratory  momentum $q_\pi$
using strategy two as explained in the text. Fitted in
each partial wave are the data between 40 and 97~MeV (filled
circles). For higher and lower energies, the phases are predicted.
The various lines refer to the contributions from the various orders 
as explained in the text.
}

\end{figure}

\newpage

\vskip 1cm

\begin{figure}[bht]
\centerline{
\epsfysize=7in
\epsffile{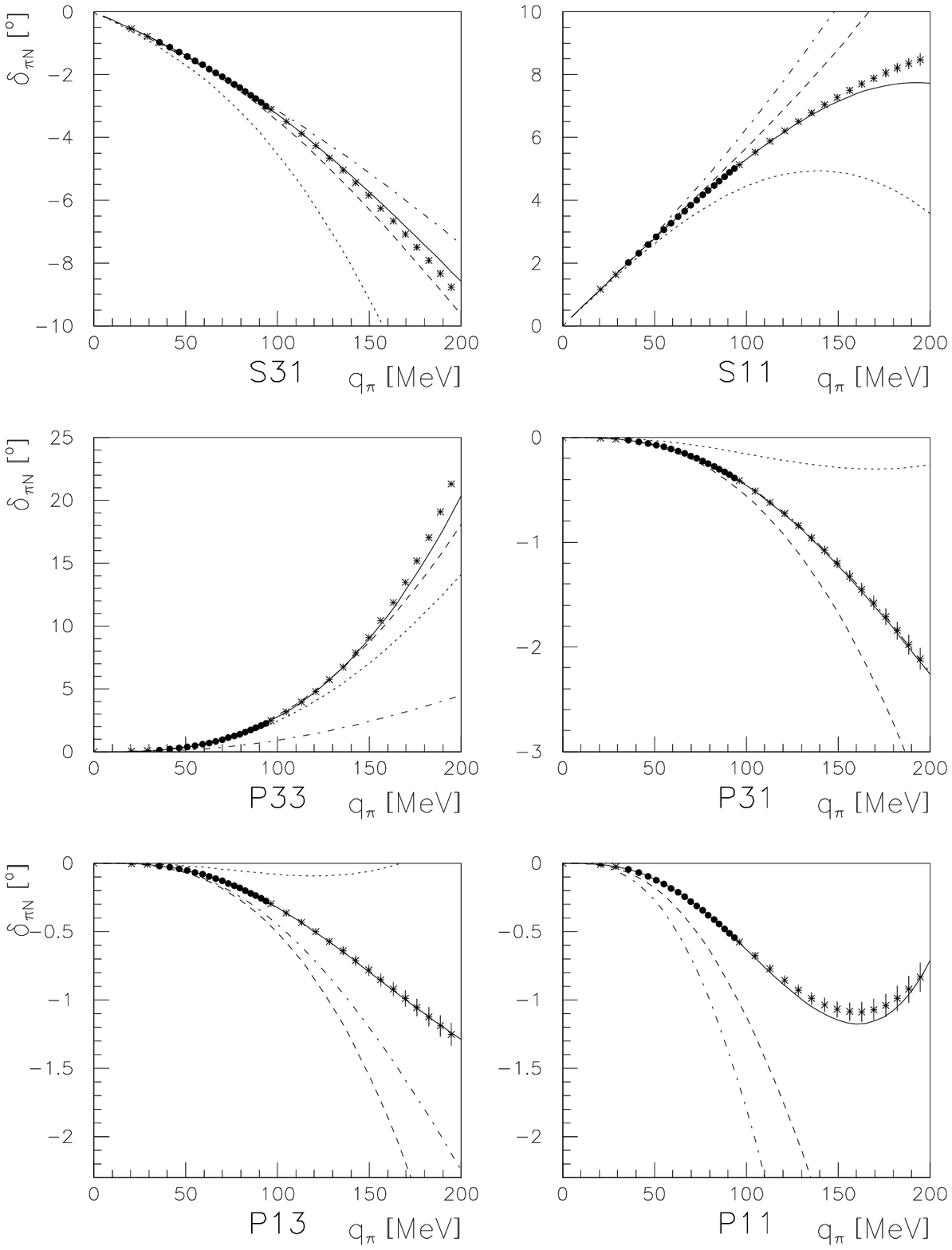}
}
\vskip 0.5cm
\caption{
Fits and predictions for the EM98 phase shifts  as a
function of $q_\pi$ using strategy two as explained in the text.  Fitted in
each partial wave are the data between 41 and 97~MeV (filled circles). For higher
and lower energies, the phases are predicted as shown by the solid lines.
The various lines refer to the contributions from the various orders 
as explained in the text.
}

\end{figure}

\newpage

\vskip 1cm

\begin{figure}[bht]
\centerline{
\epsfysize=7in
\epsffile{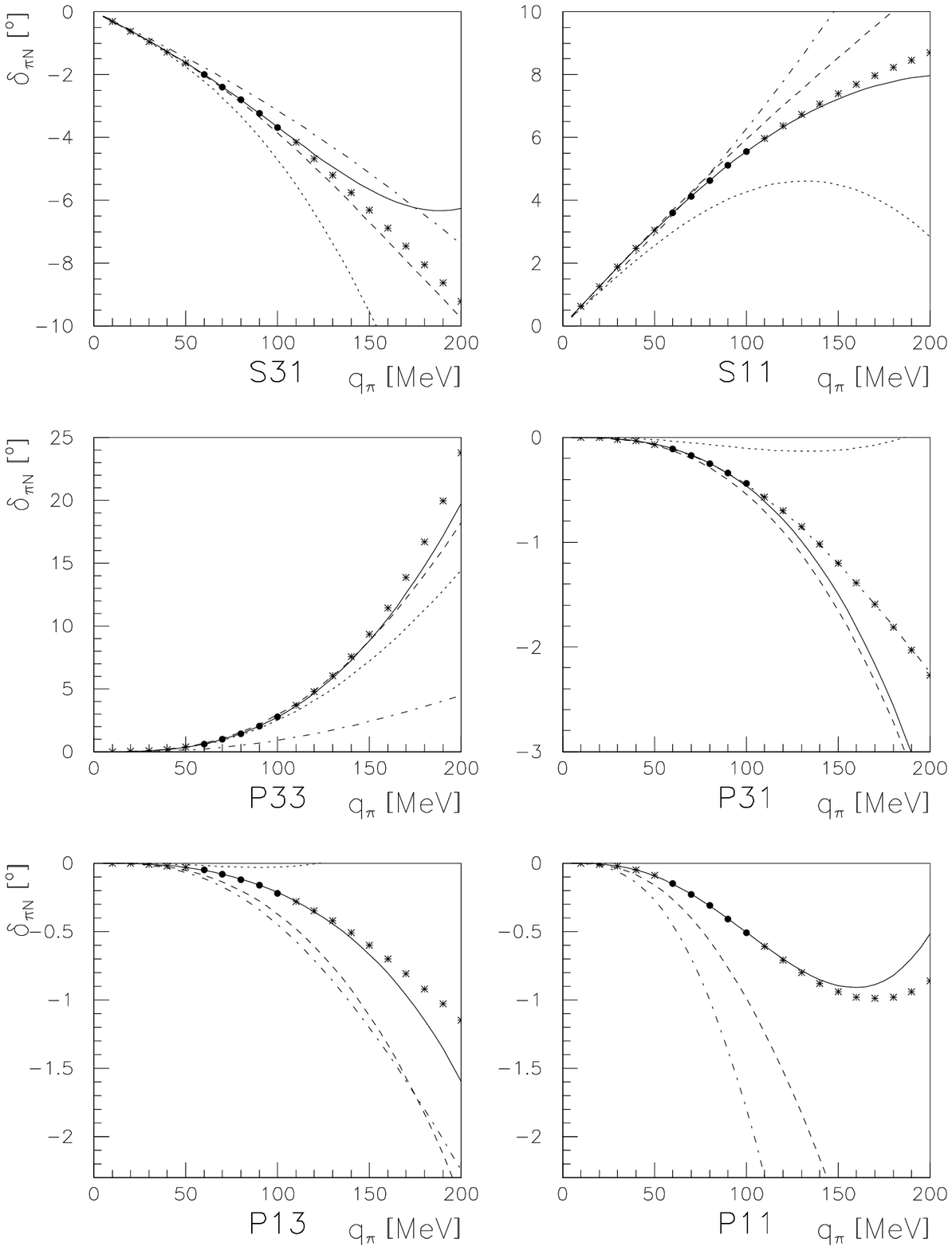}
}
\vskip 0.5cm
\caption{
Fits and predictions for the SP98 phase shifts as a
function of the pion laboratory momentum $q_\pi$
using strategy two as explained in the text. Fitted in
each partial wave are the data between 60 and 100~MeV (filled circles). For higher
and lower energies, the phases are predicted as shown by the solid
lines. The various lines refer to the contributions from the various orders 
as explained in the text.
}

\end{figure}

\end{document}